\documentclass[review]{elsarticle}

\usepackage{lineno,hyperref,amsmath,multirow,longtable,threeparttable,graphicx,epstopdf,subfig}

\modulolinenumbers[5]

\journal{arXiv.org}

\biboptions{authoryear}

\begin{document}

\begin{frontmatter}

\title{Variable Selection for High-dimensional Generalized Linear Models using an Iterated Conditional Modes/Medians Algorithm}

\author[ChulaAddress]{Vitara Pungpapong \corref{mycorrespondingauthor}}
\ead{vitara@cbs.chula.ac.th}
\cortext[mycorrespondingauthor]{Corresponding author.}
\author[PurdueAddress]{Min Zhang}
\author[PurdueAddress]{Dabao Zhang}

\address[ChulaAddress]{Department of Statistics, Faculty of Commerce and Accountancy, Chulalongkorn University, Bangkok, Thailand}
\address[PurdueAddress]{Department of Statistics, Purdue University, West Lafayette, IN, USA}

\begin{abstract}
High-dimensional linear and nonlinear models have been extensively used to identify associations between response and explanatory variables. The variable selection problem is commonly of interest in the presence of massive and complex data. An empirical Bayes model for high-dimensional generalized linear models (GLMs) is considered in this paper. The extension of the Iterated Conditional Modes/Medians (ICM/M) algorithm is proposed to build up a GLM. With the construction of pseudodata and pseudovariances based on iteratively reweighted least squares (IRLS), conditional modes are employed to obtain data-drive optimal values for hyperparameters and conditional medians are used to estimate regression coefficients. With a spike-and-slab prior for each coefficient, a conditional median can enforce variable estimation and selection at the same time. The ICM/M algorithm can also incorporate more complicated prior by taking the network structural information into account through the Ising model prior. Here we focus on two extensively used models for genomic data: binary logistic and Cox's proportional hazards models. The performance of the proposed method is demonstrated through both simulation studies and real data examples. The implementation of the ICM/M algorithm for both linear and nonlinear models can be found in the \texttt{icmm} R package which is freely available on CRAN.
\end{abstract}


\end{frontmatter}

\linenumbers

\section{Introduction}
Due to the advanced development of computing technology, high-dimensional data are widely found in many fields ranging from genomics to marketing and finance. Such data are known as large $p$ small $n$ data where the number of variables $p$ is relatively larger than the sample size $n$. Generalized linear models (GLMs) have been extensively used to identify associations between response and explanatory variables. Similar to a normal linear regression model, the classical GLMs approach also suffers in large $p$ small $n$ scenario due to collinearity among covariates. Furthermore, regression coefficients are usually assumed to be sparse. 

In biomedical science, variable selection plays an important role in analysis of high-throughput genotype. A typical microarray data have several thousands predictors. For genome-wide association studies (GWAS), the data may contain thousand to million Single nucleotide polymorphisms (SNPs). However, the sample size is usually only in hundreds or thousands. Collinearity problem occurs especially for those genes or SNPs resided in the same biological pathway. It is also believed that only few genes or SNPs are associated with the phenotype of interest. In addition to continuouse phenotype, discrete or censored outcomes frequently appear in the field of biomedical science. Binary logistic and Cox's proportional hazards models with high-dimensional covariates have received considerable attention over the past decade.

One popular approach in fitting high-dimensional GLMs is through a penalized likelihood method. The most basic and popular penalized likelihood estimator is lasso (\cite{Tibshirani1996}) where $\ell_1$-penalty is employed to produce sparse coefficients. Many works has been proposed to efficiently compute the regularization paths for GLMs with lasso penalty. \cite{Park2007} introduced the predictor-corrector method to determine the entire path of lasso for GLMs. \cite{Friedman2010} proposed a fast algorithm exploiting cyclical coordinate descent to compute a regularization path for GLMs which can shorten computing time considerably. An important extension of the lasso method is adaptive lasso (\cite{Zou2006}) which allows weights of predictors in the penalty term to be varied. \cite{Zou2006} presented that the adaptive lasso enjoys the oracle properties in linear model and GLMs under mild condition. This means that it has variable selection consistency and the ability to correctly select the nonzero coefficients with probability converging to one. Much efforts have been devoted in investigating theoretical properties of adaptive lasso. \cite{Huang2008} presented that the adaptive lasso has oracle property even when the number of covariates is much larger than the sample size in linear model. Other recent work including \cite{Wang2014} and \cite{Cui2017} also demonstrated the oracle property of the adaptive lasso in high-dimensional setting for GLMs. Another advantage of the adaptive lasso method is that it can be solved by the same efficient algorithm for solving lasso.

Alternatively, Spike-and-slab approaches to Bayesian variable selection has gained increasing attention recently. The spike part concentrates its mass at the values near zero and the slab component has a density spread over parameter spaces. There has been much work on hierarchical linear models with absolutely continuous spikes (i.e., mixture normal priors). Examples include \cite{George1993}, \cite{George1997}, \cite{Ishwaran2005}, \cite{Rockova2014}. MCMC stochastic search has been extensively used to fit the model with such Bayesian formulation in large-scale data analysis. However, it is computational expensive when the number of predictors $p$ is large. Instead of using MCMC, \cite{Rockova2014} proposed an EMVS algorithm to fit a linear model with mixture normal prior. With conjugate spike-and-slab prior, the EMVS has closed form solutions for both E- and M- steps which can save computational time over MCMC. Although an absolutely continuous spike component has ability to determine the amount of shrinkage adaptively, regression coefficients are not sparse in the exact sense. To enforce some regression coefficients to be exactly zero,  a dirac spike using a point mass at zero should be considered instead. Nonetheless, EMVS algorithm is not in the closed form anymore. Moreover, extensions of the EMVS algorithm to GLMs are not straightforward.


\cite{Pungpapong2015} presented a contribution of empirical Bayes
thresholding (\cite{Johnstone2004}) to select variables in a linear regression framework which can identify and estimate sparse predictors efficiently. A dirac spike-and-slab prior, a mixture prior of an atom of probability at zero and a heavy-tailed density, is put on each regression coefficient. An iterated conditional modes/medians (ICM/M) was also proposed for fast and easy-to-implement algorithm for empirical Bayes variable selection. Similar to an iterated conditional modes algorithm proposed by \cite{Besag1975}, conditional modes are used to obtain hyperparameters and parameters other than regression coefficients. With a dirac spike-and-slab posterior for each regression coefficient, a conditional median is employed to enforce the variable selection. As demonstrated in \cite{Pungpapong2015}, empirical Bayes variable selection can also handle the case when the information about structural relationship among predictors is available through the Ising prior. The ICM/M algorithm can be easily implemented for such complicated prior.

The aim of this paper is to generalize empirical Bayes variable selection
 to more class beyond linear model. The general framework for extension of empirical Bayes variable selection to GLMs will be described. The details on the implementation of ICM/M algorithm for binary logistic model and Cox's proportional hazards model will also be discussed here. 

The rest of of this paper is organized as follows. The next section presents a Bayesian model formulation for high-dimensional GLMs. Section 3 introduces the ICM/M algorithm for GLMs. In Section 4, we introduce how to quantify variable importance and select variables based on false discovery rate. Section 5 and Section 6 focus on the implementation of ICM/M algorithm for binary logistic model and Cox's proportional hazards model respectively. Numerical results based on simulation studies are also be shown in Section 5 and Section 6. Application to real data sets are presented in Section 7 and 8. We then conclude this paper with a discussion in Section 9.

\section{A Bayesian model formulation for high-dimensional GLMs}
Consider a generalized linear model (GLM) where each response variable $Y_i$ is assummed to be independent and has distribution belonging to exponential family taken the form
\begin{equation*}
f_{Y_i}(y_i|\theta_i, \phi) = \exp\left\{\frac{y_i\theta_i-b(\theta_i)}{a(\phi)}+c(y_i, \phi)\right\}
\end{equation*}
where $a(.)$,$b(.)$, and $c(.)$ are functions which vary according to distributions. $\theta=(\theta_i, ..., \theta_n)$ is known as the canonical parameter and $\phi$ is a dispersion parameter assummed to be known here. A link function $g(.)$ is used to connect the linear predictors $\eta_i = \beta_0+\mathbf{X}_i\beta$ to the mean $\mu_i=E[Y_i]$. That is, $g(\mu_i)=\eta_i=\beta_0+\mathbf{X}_i\beta$.

With an $n \times 1$ matrix of responses $\mathbf{Y}$ and $n \times p$ matrix containing values of $p$ predictors, the maximum likelihood estimators (MLEs) are obtained to estimate $\beta$ in low-dimensional setting ($n > p$). An iteratively reweighted least squares (IRLS) algorithm is typically used to estimate such coefficients. Based on current parameter estimates $(\hat{\beta}_0, \hat{\beta})$, a quadratic approximation to the likelihood is employed to construct pseudodata $\mathbf{Z} = (Z_1,..,Z_n)^t$ and pseudovariances $\Sigma=\text{diag}\{\sigma_i^2\}$ as follows:
\begin{align}\label{IRLSupdate}
\mathbf{Z} &= \hat{\eta}+ (\mathbf{Y}-\hat{\mu})\left(\frac{d\eta}{d\mu}\bigg|\eta=\hat{\eta}\right)\\
\Sigma &= \text{diag}\{\sigma_i^2\} = \text{diag} \left\{ Var[Y_i] \left(\frac{d\eta}{d\mu}\bigg|\eta=\hat{\eta}\right)^2\right\}
\end{align}
where $\hat{\eta} = \hat{\beta}_0+\mathbf{X}\hat{\beta}$. \cite{McCullagh1989} showed that the update of $\beta$ in IRLS is the result of a weighted least squares regression of $\mathbf{Z}$ on $\mathbf{X}$ where the weighted matrix is $\Sigma^{-1}$. Indeed, the underlying distribution of pseudodata $\mathbf{Z}$ is approximated by $N(\hat{\beta}_0+\mathbf{X}\hat{\beta}, \Sigma)$.

When $n\ll p$, a unique solution for the classical MLEs does not exist due to the fact that the design matrix $\mathbf{X}$ is not full rank. Thus, the regression coefficients $\beta$ cannot be updated using a weighted least squares in IRLS procedure. Furthermore, typical high-dimensional data analysis usually requires that only a subset of important variables are used in the modeling process. Variable selection to get sparse regression coefficients is a crucial task when analyze such massive data.

To introduce sparsity in the modeling process, an independent mixture of an atom of probability at zero and a distribution describing non-zero effect can be put on each of the regression coefficient $\beta_j$ as prior in the Bayesian framework. With the approximated distribution of pseudodata $\mathbf{Z}$ and assuming that all other parameters except $\beta_j$ are known, $\beta_j$ has a sufficient statistic $(X_j^t\Sigma^{-1}X_j)^{-1}(X_j^t\Sigma^{-1}\tilde{Z}_j)$ where $\tilde{Z}_j = \mathbf{Z} - \hat{\eta} + X_j\beta_j$ and the approximated distribution of the sufficient statistics is $N(\beta_j, (X_j^t\Sigma^{-1}X_j)^{-1})$. Following \cite{Pungpapong2015}, the prior of $\beta_j$ is in the form
\begin{equation}\label{eqn:indprior}
\beta_j | \omega \sim (1-\omega)\delta_0(\beta_j) + \omega \gamma(\beta_j).
\end{equation}
Under the prior distribution in \eqref{eqn:indprior}, each $\beta_j$ is zero with probability $(1-\omega)$ and $\beta_j$ is drawn from the nonzero part of prior $\gamma(\beta_j)$ with probability $\omega$. Laplace distribution is employed for $\gamma(\beta_j)$, that is,
\begin{equation}\label{eqn:laplace}
\gamma(\beta_j) = \frac{\alpha(X_j^t\Sigma^{-1}X_j)^{-1/2}}{2} \exp \left\{ -\alpha(X_j^t\Sigma^{-1}X_j)^{1/2} |\beta_j|\right\},
\end{equation}
where $\alpha > 0$ is a scale parameter. 

When the information of structural relationships among predictors is available, an indicator variable $\tau = (\tau_1,...,\tau_p)^t$ where $\tau_j = I_{\{\beta_j\ne0\}}$ is introduced and the underlying relationships represented by an undirected graph is put on $\tau$. Specifically, given $\tau_j$, the prior of $\beta_j$ is
\begin{equation}\label{eqn:structureprior}
\beta_j | \tau_j \sim (1-\tau_j) \delta_0(\beta_j)+\tau_j \gamma(\beta_j)
\end{equation}
Again, the Laplace distribution is employed for $\gamma(\beta_j)$. To model the relationship among $\tau$ under an undirected graph $G=(V,E)$ comprising a set $V$ of vertices and a set $E$ of edges, the following Ising model is considered:
\begin{equation}\label{eqn:isingprior}
P(\tau) = \frac{1}{Z(a,b)} \exp\{a\sum_j \tau_j + b \sum_{<j,l>\in E} \tau_j \tau_l\}
\end{equation}
where $a$ and $b$ are parameters and $Z(a,b)$ is a normalizing constant, that is, 
\begin{equation}
Z(a,b) = \sum_{\tau\in\{0,1\}^p} \exp\{a\sum_j \tau_j + b \sum_{<j,l>\in E} \tau_j \tau_l\}
\end{equation}
When $b>0$, the interaction between nearest neighboring nodes is called \textit{ferromagnetic}, i.e., neighboring $\tau_j$ and $\tau_l$ tend to have the same value. When $b<0$, the interaction is called \textit{anitiferromagnetic}, i.e., neighboring $\tau_j$ and $\tau_l$ tend to have different values. When $b=0$, the prior gets back to independent and identical Bernoulli distribution. The value of $a+b$ indicates the preferred value of each $\tau_j$. In fact, $\tau_j$ tends to be one when $a+b>0$ and $\tau_j$ tends to be zero when $a+b<0$.

\section{The iterated conditional modes/medians algorithm for GLMs}
\cite{Pungpapong2015} presented an iteratated conditional modes/medians (ICM/M) algorithm for fast computation of empirical Bayes variable selection in a linear model. Data-driven optimal values for hyperparameters and auxiliary parameters are obtained as the modes of their full conditional distribution functions. Each regression coefficient is obtained as the median of its full conditional distribution function. With a spike and slab prior on $\beta$ as in \eqref{eqn:indprior} and  \eqref{eqn:structureprior}, the regression coefficients obtained from conditional medians are sparse which therefore enforce variable selection and estimation simultaneously. The iterative procedure for updating regression coefficients and other parameters is carried out until convergence.

Adopting the idea of IRLS, we can extend the ICM/M algorithm to GLMs. Based on the approximated distribution of the current pseudo data, the ICM/M algorithm is applied to update all parameters. The procedure consists of the outer and inner loop. The outer loop is taken place to update pseudodata and pseudovariances based on current parameter estimates. The inner loop is where the ICM/M is employed to cycle through all parameters update.  

\paragraph{\textbf{Proposition 1}}  With current values of $\beta_0$ and $\beta$, pseudodata and pseudovariances $\{\mathbf{Z}, \Sigma\}$ can be calculated and a sufficient statistic for $\beta_j$ is $(X_j^t \Sigma^{-1} X_j)^{-1}(X_j^t \Sigma^{-1} \tilde{Z_j})$ w.r.t. the approximated distribution of $\mathbf{Z}$ which is $N(\beta_0+\mathbf{X}\beta, \Sigma)$. Then, the iterative conditional median of $\beta_j$ in the ICM/M algorithm can be constructed as the posterior median of $\beta_j$ in the following Bayesian analysis, 
\begin{equation*}
\left\{
\begin{array}{l}
(X_j^t \Sigma^{-1} X_j)^{-1/2}(X_j^t \Sigma^{-1} \tilde{Z}_j) | \beta_j \sim N( (X_j^t \Sigma^{-1} X_j)^{1/2}, 1)\\
\beta_j | \omega \sim (1-\omega)\delta_0(\beta_j) +\omega \frac{\alpha (X_j^t \Sigma^{-1} X_j)^{1/2}}{2} \exp \{-\alpha (X_j^t \Sigma^{-1} X_j)^{1/2} |\beta_j| \}.
\end{array}\right.
\end{equation*}
\\
With the independent prior as in \eqref{eqn:indprior}, the details of the ICM/M algorithm is demonstrated in Algorithm 1. 
\paragraph{\textbf{Algorithm 1}} 
\begin{enumerate}
\item Initialize $\left(\beta_0^{(0)}, \beta^{(0)}\right)$ and set $k=0$.
\item Set $k = k+1$.
\item Update pseudodata and pseudovariances $\left(\mathbf{Z}^{(k)}, \Sigma^{(k)}\right)$.
\item Update weight $\omega$ as the mode of its full conditional distribution function.
\begin{equation*}
\omega^{(k)} = \text{mode} P\left(\omega | \mathbf{Y}, \mathbf{X}, \mathbf{Z}^{(k)}, \Sigma^{(k)}, \beta_0^{(k)}, \beta^{(k)}\right)
\end{equation*}
\item For $j=1,...,p$, update $\beta_j$ as the posterior median. 
\begin{equation*}
\beta_j^{(k)} = \text{median} P\left(\beta_j | \mathbf{Y}, \mathbf{X}, \mathbf{Z}^{(k)}, \Sigma^{(k)}, \omega^{(k)}, \beta_0^{(k-1)}, \beta_{1:(j-1)}^{(k)}, \beta_{(j+1):p}^{(k-1)}\right)
\end{equation*}
\item Update $\beta_0$ as follows:
\begin{equation*}
\beta_0^{(k)} = \bar{Z}_w - \bar{X}_w \beta^{(k)}
\end{equation*}
where $\bar{Z}_w$ and $\bar{X}_w$ are the weighted means of $\mathbf{Z}^{(k)}$ and $\mathbf{X}$ respectively with weight for each observation being $\sigma_i^{-2}$.
\item Iterate between 2 - 6 until convergence.
\end{enumerate}

To incorporate the information of structural relationships among predictors when it is available, the Ising prior \eqref{eqn:structureprior} is employed. Since the normalizing constant of the Ising model $Z(a,b)$ is often computationally expensive, the values of hyperparameters $a$ and $b$ can be computed by maximizing the pseudo-likelihood instead of the original prior likelihood. Specifically, the pseudo-likelihood is as follows:
\begin{equation*}
L_p(a,b) = \prod\limits_{j=1}^p P \left( \tau_j | \{\tau_l: <j,l> \in E\}, a, b \right) = \prod\limits_{j=1}^p \frac{\exp \left\{ \tau_j \left(a+b\sum_{<j,l>\in E} \tau_l\right)\right\}}{1+\exp\left\{a+b\sum_{<j,l> \in E} \tau_l \right\}}
\end{equation*}
Indeed, $a$ and $b$ are the logistic regression coefficients when the binary variable $\tau_j$ is regressed on $\sum_{<j,l> \in E} \tau_l$ for $j=1,...,p$. Thus, the conditional median of $\beta_j$ can be constructed on the basis of Proposition 2 and the details of the algorithm can be found in Algorithm 2.

\paragraph{\textbf{Proposition 2}} With current values of $\beta_0$ and $\beta$, pseudodata and pseudovariances $\{\mathbf{Z}, \Sigma\}$ can be calculated and a sufficient statistic for $\beta_j$ is $(X_j^t \Sigma^{-1} X_j)^{-1}(X_j^t \Sigma^{-1} \tilde{Z_j})$ w.r.t. the approximated distribution of $\mathbf{Z}$ which is $N(\beta_0+\mathbf{X}\beta, \Sigma)$. Then, the iterative conditional median of $\beta_j$ in the ICM/M algorithm can be constructed as the posterior median of $\beta_j$ in the following Bayesian analysis, 
\begin{equation*}
\left\{
\begin{array}{l}
(X_j^t \Sigma^{-1} X_j)^{-1/2}(X_j^t \Sigma^{-1} \tilde{Z}_j) | \beta_j \sim N( (X_j^t \Sigma^{-1} X_j)^{1/2}, 1)\\
\beta_j | \varpi_j \sim (1-\varpi_j)\delta_0(\beta_j) +\varpi_j \frac{\alpha (X_j^t \Sigma^{-1} X_j)^{1/2}}{2} \exp \{-\alpha (X_j^t \Sigma^{-1} X_j)^{1/2} |\beta_j| \},
\end{array}\right.
\end{equation*}
where the probability $\varpi_j$ is specified as follows,
\begin{equation*}
\varpi_j^{-1} = 1+\exp\left\{-a-b \sum\limits_{l:<j,l> \in E} \tau_l\right\}
\end{equation*}

\paragraph{\textbf{Algorithm 2}} 
\begin{enumerate}
\item Initialize $\left(\beta_0^{(0)}, \beta^{(0)}\right)$ and set $k=0$.
\item Set $k = k+1$.
\item Update pseudodata and pseudovariances $\left(\mathbf{Z}^{(k)}, \Sigma^{(k)}\right)$.
\item Update $\tau^{(k)} = (\tau_1^{(k)}, ..., \tau_p^{(k)})$ where $\tau_j^{(k)} = I_{\{ \beta_j^{(k-1)} \ne 0\}}$.
\item Update hyperparameters $(a,b)$ as the mode of its pseudo-likelihood function.
\begin{equation*}
\left(a^{(k)}, b^{(k)} \right) = \text{mode} \left\{\prod_{j=1}^p P \left(\tau_j^{(k)} | \tau_{l}^{(k)} : <j,l> \in E \right) \right\}
\end{equation*}
\item For $j=1,...,p$, update $\beta_j$ as the posterior median. 
\begin{align*}
\beta_j^{(k)} &= \text{median} P\left(\beta_j | \mathbf{Y}, \mathbf{X}, \mathbf{Z}^{(k)}, \Sigma^{(k)}, \tau^{(k)}, a^{(k)}, b^{(k)} \beta_0^{(k-1)}, \beta_{1:(j-1)}^{(k)}, \beta_{(j+1):p}^{(k-1)}\right) \\
&= \text{median} P\left(\beta_j | \mathbf{Y}, \mathbf{X}, \mathbf{Z}^{(k)}, \Sigma^{(k)}, \varpi_j^{(k)}, \beta_0^{(k-1)}, \beta_{1:(j-1)}^{(k)}, \beta_{(j+1):p}^{(k-1)}\right)
\end{align*}
where the probability $\varpi_j^{(k)} = \left [ 1+\exp\left\{-a^{(k)}-b^{(k)} \sum \limits_{l:<j,l>\in E} \tau_l^{(k)} \right\} \right ]^{-1}$.
\item Update $\beta_0$ as follows:
\begin{equation*}
\beta_0^{(k)} = \bar{Z}_w - \bar{X}_w \beta^{(k)}
\end{equation*}
where $\bar{Z}_w$ and $\bar{X}_w$ are the weighted means of $\mathbf{Z}^{(k)}$ and $\mathbf{X}$ respectively with weight for each observation being $\sigma_i^{-2}$.
\item Iterate between 2 - 7 until convergence.
\end{enumerate}

\section{Variable importance and false discovery rate}
Although the ICM/M algorithm with a spike-and-slab prior can enforce variable estimation and selection simultaneously, it is often of interest to quantify the importance of variables. \cite{Pungpapong2015} proposed a local posterior probability to evaluate the importance of variables. Specifically, given all other parameter estimates except for $\beta_j$, a local posterior probability of $j$-th predictor is defined as
\begin{align} \label{eqn:localpostprob}
\begin{split}
\zeta_j &= P(\beta_j \ne 0 | \mathbf{Y}, \mathbf{X}, \hat{\beta}_0, \hat{\beta}_{1:(j-1)}, \hat{\beta}_{(j+1):p}, .) \\
&= \int P(\beta_j | \mathbf{Y}, \mathbf{X}, \hat{\beta}_0, \hat{\beta}_{1:(j-1)}, \hat{\beta}_{(j+1):p}, .).
\end{split}
\end{align}
However, with the Bayesian formulation for GLMs, the local posterior probability in \eqref{eqn:localpostprob} is not in the closed form and it is natural to estimate it by using the pseudodata in the last iteration $(\mathbf{Z}, \Sigma)$ and its approximated distribution. That is, 
\begin{equation}\label{eqn:approxlocalpostprob}
\zeta_j \approx \int P(\beta_j | \mathbf{Z}, \Sigma, \mathbf{X}, \hat{\beta}_0, \hat{\beta}_{1:(j-1)}, \hat{\beta}_{(j+1):p}, .).
\end{equation}
Such probability in \eqref{eqn:approxlocalpostprob} has a closed form and can be easily computed. 

With the local posterior probability $\zeta$ and true $\beta$, a true false discovery rate (FDR) given the data can be computed as
\begin{equation}\label{eqn-trueFDR}
FDR = \frac{\displaystyle\sum\limits_{j=1}^p 1_{\{\beta_j=0\}} 1_{\{\zeta_j>\kappa\}} }{\displaystyle\sum\limits_{j=1}^p
1_{\{\zeta_j>\kappa\}}}, \ \ \ \ \ 0 \le \kappa < 1
\end{equation}
Following \cite{Newton2004}, the expected FDR given the data in Bayesian scheme is defined as
\begin{equation}\label{eqn-estFDR}
\widehat{FDR} = \frac{\displaystyle\sum\limits_{j=1}^p (1-\zeta_j) 1_{\{\zeta_j>\kappa\}}}{\displaystyle\sum\limits_{j=1}^p
1_{\{\zeta_j>\kappa\}}}
\end{equation}
We then can select a set of important predictors based on the local posterior probability and $\widehat{FDR}$. By controlling $\widehat{FDR}$ at prespecified level, $\kappa$ can be chosen and a set of important predictors lists all predictors having the local posterior probability greater than $\kappa$. It has been shown that the local posterior probability is a good indicator to quantify the importance of variables.

\section{The ICM/M algorithm for binary logistic regression}
\subsection{Impplementation details}
When the response variable is binary taking values $\{0,1\}$, the logistic regression is commonly used. The logistic regression assumes $Y_i$ is independently Bernoulli distributed with mean $E[Y_i] = P(Y_i=1) = \pi(\mathbf{X}_i)$ and variance $Var[Y_i] = \pi(\mathbf{X}_i)(1-\pi(\mathbf{X}_i))$ where
\begin{equation}\label{eqn:logisticreg}
\pi(\mathbf{X}_i) = \frac{e^{\beta_0+\mathbf{X}_i\beta}}{1+e^{\beta0+\mathbf{X}_i\beta}}.
\end{equation}
Equivalently, the logistic regression model can be written as
\begin{equation}
\log\left( \frac{\pi(\mathbf{X}_i)}{1-\pi(\mathbf{X}_i)} \right) = \beta_0+\mathbf{X}_i\beta.
\end{equation}

The pseudodata and pseudovariance for binary logistic regression can be constructed based on the current parameter estimates $(\hat{\beta}_0, \hat{\beta})$ as
\begin{align}
Z_i &= \hat{\beta_0}+\mathbf{X}_i\hat{\beta} + \frac{Y_i -\hat{\pi}(\mathbf{X}_i)}{\hat{\pi}(\mathbf{X}_i)(1-\hat{\pi}(\mathbf{X}_i))}\\
\sigma_i^2 &= \frac{1}{\hat{\pi}(\mathbf{X}_i)(1-\hat{\pi}(\mathbf{X}_i))}
\end{align}

where $\hat{\pi}(\mathbf{X}_i) = \frac{e^{\hat{\beta}_0+\mathbf{X}_i\hat{\beta}}}{1+e^{\hat{\beta}_0+\mathbf{X}_i\hat{\beta}}}$ for $i=1,...,n$. 

Some cautions need to be considered to avoid any divergence issues. The details in the implementation of ICM/M algorithm are discussed here to prevent divergence problem:
\begin{itemize}
\item When a probability is within the range of $\epsilon =10^{-5}$ of 0 or 1, we set it to 0 and 1 respectively.
\item When a probability is close to 1. Numeric difficulty arises in calculating pseudodata and pseudovariances due to the term $\hat{\pi}(\mathbf{X}_i)(1-\hat{\pi}(\mathbf{X}_i))$ in the denominator. Hazard or tail functions are employed to improve numerical stability in IRLS (\cite{Jorgensen1994}). We define
\begin{equation*}
F(\eta_i) = \frac{e^{\eta_i}}{1+e^{\eta_i}} = \frac{1}{1+e^{-\eta_i}},
\end{equation*}
\begin{equation*}
h_{+}(\eta_i) = F(\eta_i) \;\;\;\;\;\; \text{and} \;\;\;\;\; h_{-}(\eta_i) = 1 - F(\eta_i).
\end{equation*}
The pseudodata and pseudovariance can be computed as
\begin{align*}
Z_i &= \left \{ 
           \begin{array}{ll}
            \hat{\eta}_i + \frac{1}{h_{+}(\hat{\eta}_i)} & \text{, if } Y_i = 1 \\
            \hat{\eta}_i - \frac{1}{h_{-}(\hat{\eta}_i)} & \text{, if } Y_i = 0
           \end{array} \right . \\
\sigma^2_i &= \frac{1}{h_{-}(\hat{\eta}_i)h_{+}(\hat{\eta}_i)}.
\end{align*}
In addition, to avoid the overflows, when $\hat{\eta}_i < -30$ and $\hat{\eta}_i>30$, $\hat{\eta}_i$ is set to -30 and 30 respectively. Care should also be taken to select the appropriate form of the function in each tail. Specifically, the form involving $e^{-\eta}$ should be used for positive value of $\eta$ and the form involving $e^{\eta}$ should be used for negative value of $\eta$.
\item The idea of active-set convergence is adopted to check convergence in each iteration. The algorithm stops when a complete cycle to update the coefficients in the inner loop does not change the active set of predictors - those with nonzero coefficients. Active set convergence is also mentioned in \cite{Friedman2010}, \cite{Meier2008}, and \cite{Krishnapuram2005}.
\end{itemize}

\subsection{Simulation studies}
Simulation studies were conducted to evaluate the performance of our proposed logistic empirical Bayes variable selection via the ICM/M algorithm. Large $p$ small $n$ data sets were simulated from the model \eqref{eqn:logisticreg}. We compare the performance of our approach with other two popular methods: the regularized logistic regression with lasso (Lasso) and adaptive lasso (ALasso) penalties. A tuning parameter for both methods was chosen based on 10-fold cross-validation. Using lasso fits as initial values, the ICM/M algorithm was carried out. 

Three cases of prior on structured predictors were considered here. Case 1 assumed that all predictors are mutually independent. For Case 2 and Case 3, the information of structural relationships among predictors was assumed to be known and the ICM/M with the Ising prior was applied in these two cases. Average misclassification rates (MR) calculated on the test data sets, number of false positives (FP), number of false negatives (FN), model sizes (MS), ${\parallel}{\beta}{-}{\hat{\beta}}{\parallel}_1 = \sum_{j-1}^p |\beta_j-\hat{\beta}_j|$, and ${\parallel}{\beta}{-}{\hat{\beta}}{\parallel}_2^2=\sum_{j=1}^p |\beta_j-\hat{\beta}_j|^2$ were calculated among 100 simulated data sets. 

\subsubsection{Case 1: Independent prior}
The data for case 1 was generated with $n=250$ and $p=1,000$. The intercept term $\beta_0$ was set to zero and there are 10 non-zero regression coefficients including  $\beta_1=...=\beta_5=10$ and $\beta_{11}=...=\beta_{15}=-5$. The covariates were partitioned into 10 blocks, where each block containing 100 covariates were serially correlated at the same level $\rho$. The values of $\rho$ were $\{0, 0.3, 0.5, 0.7, 0.9\}$.

\begin{table}[htbp] \footnotesize
	\centering 
	\caption{Average misclassification rates (MR) calculated on the test data sets, number of false positives (FP), number of false negatives (FN), model sizes (MS), ${\parallel}{\beta}{-}{\hat{\beta}}{\parallel}_1$, and ${\parallel}{\beta}{-}{\hat{\beta}}{\parallel}_2^2$ across 100 simulated data sets in Case 1 (with standard deviation in parentheses) .} \label{table:case1-result}
	\begin{tabular}{ l | c c c c c c} 
		\hline \hline
		\multicolumn{7}{c}{$\rho = 0$} \\
		\hline 
		Method & MR & FP & FN & MS &  ${\parallel}{\beta}{-}{\hat{\beta}}{\parallel}_1$ & ${\parallel}{\beta}{-}{\hat{\beta}}{\parallel}_2^2$\\
		\hline
		Lasso & .22(.03) & 74.49(15.38) & 3.57(1.60) & 90.92(15.95) & 29.53(1.30) & 32.48(3.59) \\
		ALasso & .22(.04) & 63.21(9.57) & 3.90(1.60) & 79.31(9.61) & 29.12(2.65) & 22.83(3.99) \\
		ICM/M & .20(.04) & 6.79(3.58) & 6.73(1.60) & 20.06(3.73) & 19.06(2.86) & 16.78(3.68) \\
		\hline
		\multicolumn{7}{c}{$\rho = 0.3$} \\
		\hline
		Lasso & .16(.03) & 68.49(13.43) & 2.35(1.25) & 86.14(13.64) & 28.09(1.19) & 29.50(3.60)\\
		ALasso & .15(.03) & 43.74(8.06) & 2.67(1.46) & 61.07(8.17) & 24.43(2.43) & 19.80(3.15) \\
		ICM/M & .14(.03) & 4.87(3.21) & 5.40(1.44) & 19.47(3.48) & 16.58(2.13) & 14.34(2.67)\\
		\hline
		\multicolumn{7}{c}{$\rho = 0.5$} \\
		\hline
		Lasso & .13(.02) & 60.67(12.02) & 2.34(1.30) & 78.33(12.57) & 27.26(12.57) & 29.04(3.03) \\
		ALasso & .11(.02) & 31.52(7.33) & 2.47(1.40) & 49.05(7.83) & 22.10(1.96) & 19.37(3.25) \\
		ICM/M & .12(.03) & 3.71(2.35) & 5.55(1.39) & 18.16(2.96) & 16.45(2.11) & 15.53(3.11)\\
		\hline
		\multicolumn{7}{c}{$\rho = 0.7$} \\
		\hline
		Lasso & .10(.02) & 53.86(10.67) & 3.62(1.36) & 70.24(10.89) & 26.48(1.03) & 28.20(3.04) \\
		ALasso & .08(.02) & 17.69(4.65) & 3.42(1.43) & 34.27(4.93) & 20.19(1.57) & 20.09(2.67) \\
		ICM/M & .09(.02) & 3.85(2.70) & 6.49(1.31) & 17.36(3.21) & 17.98(1.87) & 18.71(2.75)\\
		\hline
		\multicolumn{7}{c}{$\rho = 0.9$} \\
		\hline
		Lasso & .07(.02) & 40.13(9.19) & 6.71(1.57) & 53.42(9.32) & 25.78(1.76) & 29.06(2.93)\\
		ALasso & .05(.01) & 6.84(2.90) & 6.30(1.56) & 20.54(3.43) & 20.22(1.34) & 24.49(2.81)\\
		ICM/M & .07(.02) & 3.91(2.37) & 9.03(1.49) & 14.88(3.01) & 21.19(2.15) & 25.80(3.61)\\
		\hline
	\end{tabular}
\end{table}

From Table \ref{table:case1-result}, misclassification rates among the three methods were comparable and they tended to get smaller when the correlation among predictors was high.

In order to assess the ability to select the correct variables, average number of false positives and false negatives are also reported in Table 1. Lasso produced large number of false positives across all values of $\rho$. Adaptive lasso had relatively lower false positive rates than lasso. ICM/M had the smallest number of false positive rates among the three methods. However, it had relatively higher number of false negatives. Furthermore, we also observed that the number of false positives got smaller and the number of false negatives got bigger for higher value of $\rho$. While the true model contained only 10 non-zero regression coefficients, the average model size from lasso was five to nine times larger than those in the true model. Adaptive lasso selected moderate numbers of non-zero regression coefficients. The average model sizes of ICM/M was closest to the true model. However, several true positives are missed especially when the correlation among predictors was high $(\rho = 0.9)$. 

For estimation errors, ICM/M outperformed the other two methods in terms of both ${\parallel}{\beta}{-}{\hat{\beta}}{\parallel}_1$ and ${\parallel}{\beta}{-}{\hat{\beta}}{\parallel}_2^2$ except when $\rho=0.9$ that adaptive lasso was better than ICM/M.

\begin{figure}[htp]
	\captionsetup[subfloat]{farskip=0pt,captionskip=0.1pt}
	\centering
	\subfloat[$\rho = 0$]{{\includegraphics[scale=0.65]{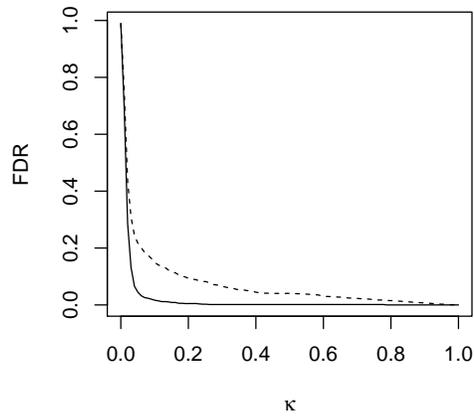} }} 
	\subfloat[$\rho=0.5$]{{\includegraphics[scale=0.65]{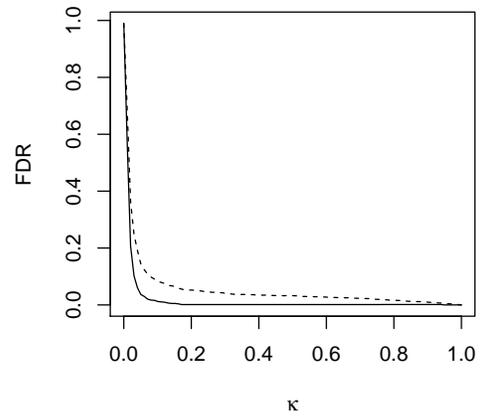} }}%
\\[-0.8cm]
	\subfloat[$\rho=0.9$]{{\includegraphics[scale=0.65]{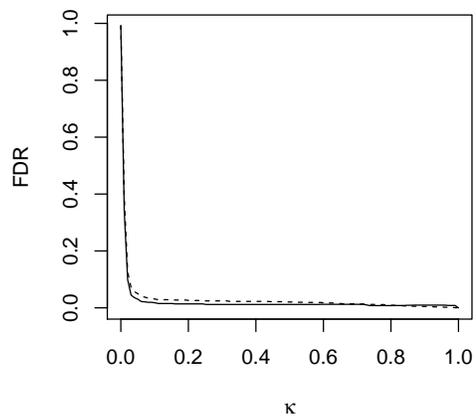} }}%
	\caption{$\kappa$ versus average true FDR (solid line) and estimated FDR (dash line) among 100 simulations in Case 1.}%
	\label{figure:case1-FDRplot}
\end{figure}

In order to evaluate the importance of variables through the local posterior probability obtained from ICM/M, the true and estimated FDR were plot against $\kappa$ in Figure \ref{figure:case1-FDRplot} . We first noticed that the estimated FDR was inflated from the true FDR. Furthermore, the FDR was well approximated when the correlation among covariates got higher. By controlling the FDR at level 0.05, the threshold of $\kappa$ was chosen and all variables having local posterior probabilities suppassed the threshold were selected. Table \ref{table:case1-kappa} shows the results on the value of $\kappa$, number of false positives, number of false negatives, and number of important variables for Case 1 simulation. We observed that the value of $\kappa$ was decreasing for higher value of $\rho$ causing lower number of false positives and higher number of false negatives. Overall, the results in Table \ref{table:case1-kappa} are better than those in Table \ref{table:case1-result}.  Thus, the local posterior probability and estimated FDR was a suitable tool to select important variables.

\begin{table}[htbp]\footnotesize
	\centering
	\caption{Average $\kappa$, number of false positives (FP), number of false negatives (FN), and model sizes (MS) across 100 simulated data sets in Case 1 when controlling estimated FDR at level 0.05 (with standard deviation in parentheses) .} \label{table:case1-kappa}
	\begin{tabular}{ c | c c c c}
		\hline \hline
		$\rho$ & $\kappa$ & FP & FN & No. of important variables\\
		\hline
		0 & 0.3876(0.2217) & 0.01(0.10) & 2.99(1.06) & 7.02(1.05)\\
		0.3 & 0.2690(0.2385) & 0.00(0.00) & 2.88(0.84) & 7.12(0.84)\\
		0.5 & 0.2679(0.2799) & 0.01(0.10) & 3.65(1.22) & 6.36(1.23) \\
		0.7 & 0.2278(0.2740) & 0.02(0.14) & 5.10(1.12) & 4.92(1.11)\\
		0.9 & 0.1430(0.2512) & 0.04(0.20) & 5.81(0.92) & 4.23(0.93)\\
		\hline
	\end{tabular}
\end{table}

\subsubsection{Case 2: Linear chain prior}
For case 2, we again fixed $n=250$ and $p=1000$. The intercept term $\beta_0$ was set to zero and the values of regression coefficients $\beta$ depend on an indicator variable $\tau$ which follows a Markov linear chain model with transition matrix:
\begin{equation*}
T = \bordermatrix{& \tau_{j+1}=0 & \tau_{j+1}=1 \cr
                \tau_j=0 & 0.99 & 0.01 \cr
                \tau_j=1 & 0.5 & 0.5 }
\end{equation*}
And $\tau_1 \sim \pi$, where $\pi = (0.5 \;\;\;  0.5)$. The effect size for non-zero coefficients were drawn from Uniform[3,10]. The covariates $\mathbf{X}$ were generated from $AR(1)$ with different value of $\rho$ in $\{0, 0.3, 0.5, 0.7, 0.9\}$.

\begin{table}[htbp]\footnotesize
\centering
\caption{Average misclassification rates (MR) calculated on the test data sets, number of false positives (FP), number of false negatives (FN), model sizes (MS), ${\parallel}{\beta}{-}{\hat{\beta}}{\parallel}_1$, and ${\parallel}{\beta}{-}{\hat{\beta}}{\parallel}_2^2$ across 100 simulated data sets in Case 2 (with standard deviation in parentheses) .} \label{table:case2-result}
\begin{tabular}{ l | c c c c c c}
\hline \hline
\multicolumn{7}{c}{$\rho = 0$} \\
\hline 
Method & MR & FP & FN & MS &  ${\parallel}{\beta}{-}{\hat{\beta}}{\parallel}_1$ & ${\parallel}{\beta}{-}{\hat{\beta}}{\parallel}_2^2$\\
\hline
Lasso  & .42(.13) & 20.57(34.61) & 15.84(8.55) & 25.71(43.04) & 161.39(0.89) & 1267.96(51.56)\\
ALasso & .20(.04) & 62.27(8.71) & 2.16(1.32) & 81.11(8.99) & 158.71(2.87) & 1093.19(36.97)\\
ICM/M & .17(.05) & 6.68(4.56) & 4.79(3.33) & 22.89(6.82) & 153.17(2.39) & 1139.08(35.42)\\
\hline
\multicolumn{7}{c}{$\rho = 0.3$} \\
\hline
Lasso & .31(.16) & 44.61(37.94) & 9.42(9.54) & 56.18(47.25) & 160.71(1.38) & 1227.51(61.36)\\
ALasso & .18(.03) & 54.86(8.75) & 2.32(1.24) & 73.54(8.90) & 157.12(2.53) & 1090.13(30.31)\\
ICM/M & .18(.06) & 3.91(4.36) & 7.33(3.59) & 17.58(7.10) & 154.04(2.31) & 1160.13(37.45)\\
\hline
\multicolumn{7}{c}{$\rho = 0.5$} \\
\hline
Lasso &  .21(.11) & 63.87(26.07) & 4.39(6.28) & 80.47(31.87) & 160.30(1.19) & 1195.71(44.00)\\
ALasso & .17(.03) & 51.59(7.59) & 2.72(1.41) & 69.87(7.85) & 157.36(2.36) & 1099.11(30.94)\\
ICM/M & .20(.03) & 2.52(3.68) & 10.13(2.09) & 13.39(4.88) & 155.42(1.46) & 1184.00(23.11)\\
\hline
\multicolumn{7}{c}{$\rho = 0.7$} \\
\hline
Lasso & .16(.04) & 67.91(13.16) & 2.72(2.34) & 86.19(14.18) & 160.33(1.54) & 1178.15(29.07)\\
ALasso & .15(.03) & 44.04(6.84) & 3.04(1.54) & 62.00(6.85) & 156.90(2.49) & 1107.57(29.79)\\
ICM/M & .18(.03) & 2.80(1.94) & 11.73(1.97) & 12.07(2.23) & 156.09(1.71) & 1191.41(20.13)\\
\hline
\multicolumn{7}{c}{$\rho = 0.9$} \\
\hline
Lasso & .13(.02) & 60.65(7.98) & 6.27(2.13) & 75.38(7.72) & 162.85(2.37) & 1183.95(24.90)\\
ALasso & .12(.03) & 35.02(5.70) & 6.70(2.21) & 49.32(5.99) & 160.01(2.89) & 1150.78(28.78)\\
ICM/M & .15(.03) & 4.83(1.86) & 14.83(1.65) & 11.00(1.80) & 158.63(2.05) & 1209.87(19.60)\\
\hline
\end{tabular}
\end{table}

As shown in Table \ref{table:case2-result}, ICM/M had smallest misclassification rates when predictors were independent or had mild correlation among them while adaptive lasso had smallest misclassification rates for mild/high correlation among predictors. Again, we saw the same pattern as in Case 1 simulation that misclassification rates were lower when correlation among predictors was higher for all three methods. 

ICM/M had much smaller number of false positives comparing to lasso and adaptive lasso across values of $\rho$. Lasso and adaptive lasso generally selected large number of non-zero predictors as the true model in Case 2 simulation contains 21 non-zero coefficients. ICM/M produced relatively larger number of false negatives and smaller model sizes especially for high correlation among predictors.

When comparing estimation errors, ICM/M yielded the smallest ${\parallel}{\beta}{-}{\hat{\beta}}{\parallel}_1$ while adaptive lasso yielded the smallest ${\parallel}{\beta}{-}{\hat{\beta}}{\parallel}_2^2$ across all values of $\rho$. Thus, it depended on the circumstances to choose between ICM/M and adaptive lasso based on estimation errors. If we would like to avoid large errors, adaptive lasso is preferred since it penalized large deviation more heavily. But if we prefer a model that the estimates are generally close to the actual values but miss badly in some of the coefficients then ICM/M is more appropriate than adaptive lasso.

\begin{figure}[htp]
	\captionsetup[subfloat]{farskip=0pt,captionskip=0.1pt}
	\centering
	\subfloat[$\rho = 0$]{{\includegraphics[scale=0.65]{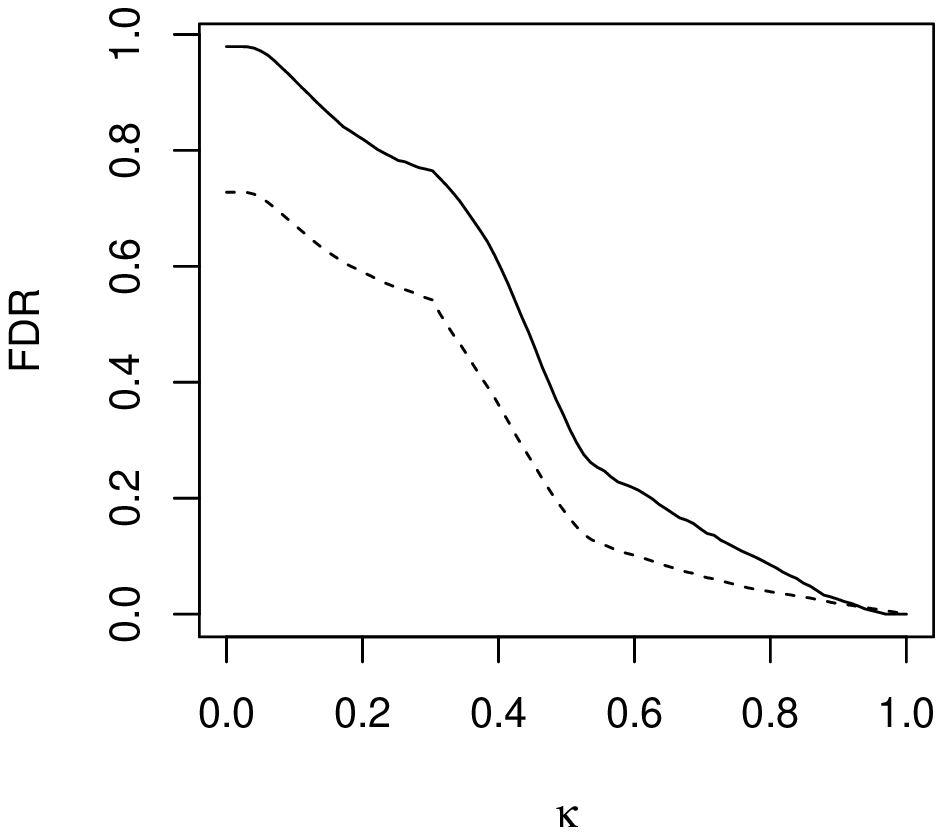} }} 
	\subfloat[$\rho=0.5$]{{\includegraphics[scale=0.65]{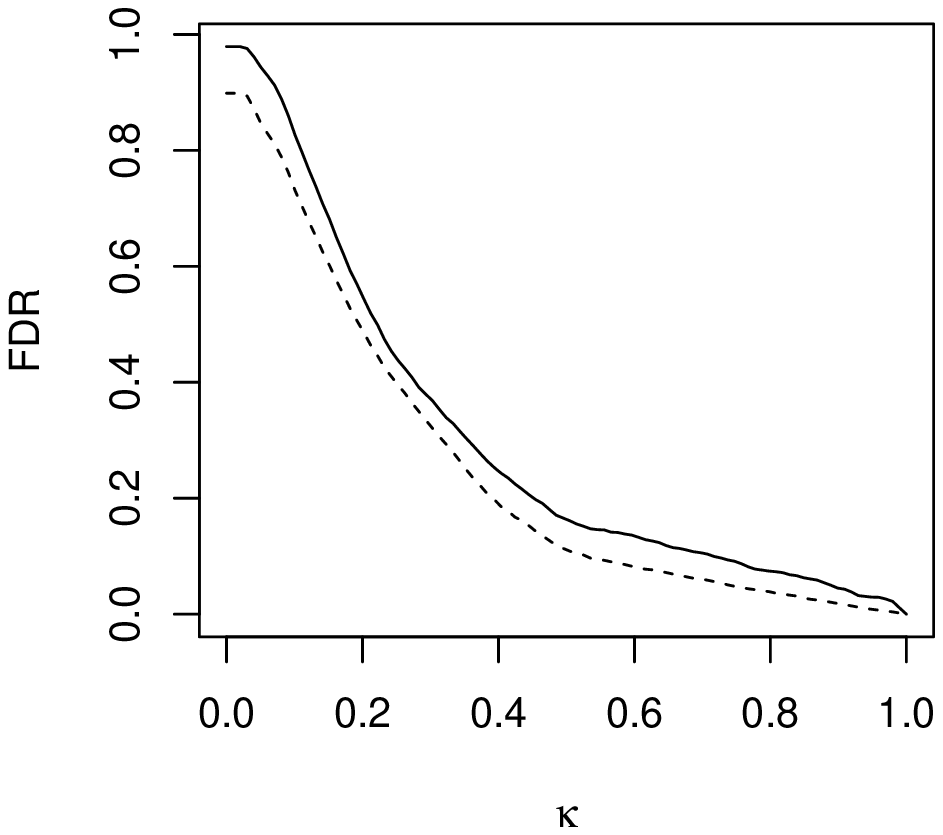} }}%
	\\[-0.8cm]
	\subfloat[$\rho=0.9$]{{\includegraphics[scale=0.65]{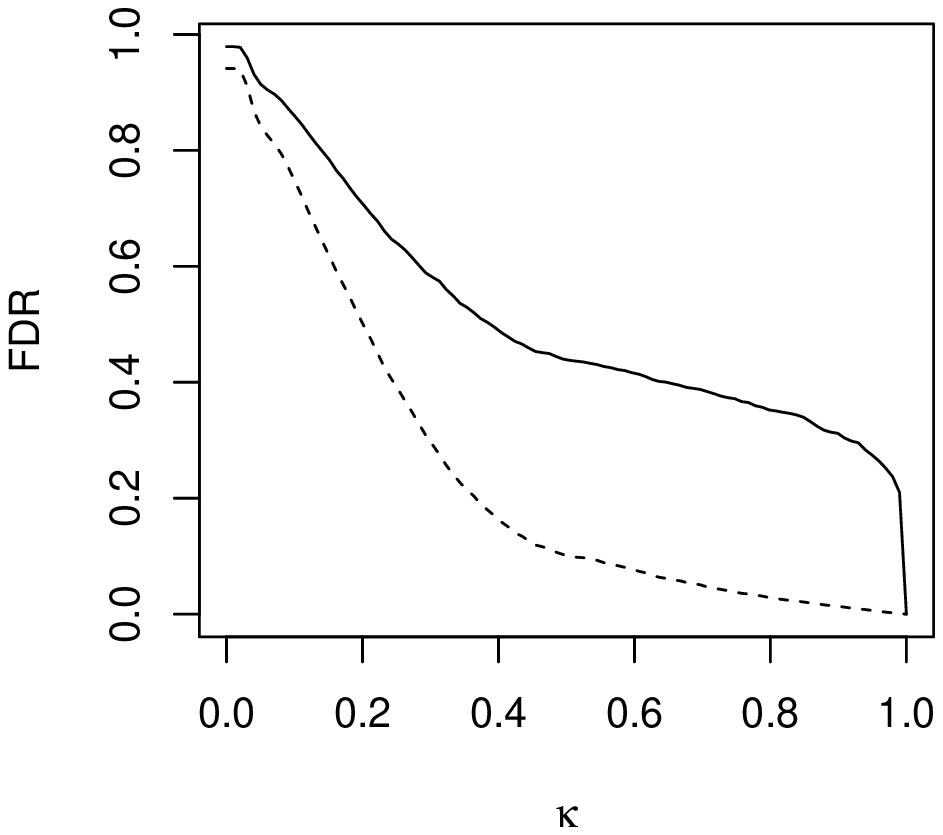} }}%
	\caption{$\kappa$ versus average true FDR (solid line) and estimated FDR (dash line) among 100 simulations in Case 2.}%
	\label{figure:case2-FDRplot}
\end{figure}

With the Ising prior, the estimated FDR curves obtained from ICM/M were underestimated making the chosen value of $\kappa$ to be smaller than it supposed to be. When FDR was controlled at level 0.05, the results in Table \ref{table:case2-kappa} had smaller number of false positives and slightly higher number of false negatives. To reduce the number of false negatives, we can increase the FDR level.

\begin{table}[htbp]\footnotesize
	\centering
	\caption{Average $\kappa$, number of false positives (FP), number of false negatives (FN), and model sizes (MS) across 100 simulated data sets in Case 2 when controlling estimated FDR at level 0.05 (with standard deviation in parentheses) .} \label{table:case2-kappa}
	\begin{tabular}{ c | c c c c}
		\hline \hline
		$\rho$ & $\kappa$ & FP & FN & No. of important variables \\
		\hline
		0 & 0.7506(0.0747) & 1.66(1.49) & 6.69(3.85) & 15.97(4.53) \\
		0.3 & 0.7132(0.1092) & 1.00(1.37) & 9.17(3.86) & 12.83(4.49)\\
		0.5 & 0.7142(0.1134) & 0.92(1.47) & 11.8(2.03) & 10.12(2.82)\\
		0.7 & 0.7192(0.0963) & 1.36(1.21) & 13.02(1.93) & 9.34(1.93)\\
		0.9 & 0.6757(0.1173) & 3.22(1.59) & 15.53(1.41) & 8.69(1.70)\\
		\hline
	\end{tabular}
\end{table}

\subsubsection{Case 3: Pathway structure prior}
 To assess the performance of our proposed method in pathway-based analysis, the data was simulated based on the genotype from publicly available Parkinson's disease (PD) dataset (dbGaP study accession number: phs000089.v3.p2). Parkinson metabolic pathway and other six pathways related to PD were obtained from KEGG database. For each genetic region in these pathways, SNPs resided in the genetic region were mapped to the gene based on their physical locations. With this procedure, most genes had at least one affiliated SNPs and there were few genes with more than thousand associated SNPs. To make the simulated data manageable, we selected all SNPs for genes having 5 or less than 5 associated SNPs. For those genes having more than 5 associated SNPs, only 5 SNPs representing each gene were randomly chosen. This process yielded $p=1,152$ SNPs representing 341 genes left for analysis. The phenotype $\mathbf{Y}$ was generated using logistic regression with these 1,152 SNPs as covariates. The regression coefficients for all 46 SNPs affiliated to 15 genes resided in Parkinson metabolic pathway were chosen to have non-zero effect size. The location of non-zero coefficients were fixed but the effect sizes were random and drawn from Uniform[1,10]. There were 1,741 individuals in the PD dataset which we randomly selected 871 individuals to be served as training data and 870 individuals as test data. The simulation was run for 100 times. 

\begin{table}[htbp]\footnotesize
\centering
\caption{Average misclassification rates (MR) calculated on the test data sets, number of false positives (FP), number of false negatives (FN), model sizes (MS), ${\parallel}{\beta}{-}{\hat{\beta}}{\parallel}_1$, and ${\parallel}{\beta}{-}{\hat{\beta}}{\parallel}_2^2$ across 100 simulated data sets in Case 3 (with standard deviation in parentheses) .} \label{table:case3-result}
\begin{tabular}{ l | c c c c c c}
	\hline \hline
	Method & MR & FP & FN & MS &  ${\parallel}{\beta}{-}{\hat{\beta}}{\parallel}_1$ & ${\parallel}{\beta}{-}{\hat{\beta}}{\parallel}_2^2$\\
	\hline
	Lasso & .10(.01) & 165.67(21.81) & 9.14(2.43) & 202.53(22.57) & 198.11(14.11) & 935.06(124.54)\\
	ALasso & .08(.01) & 47.24(13.25) & 11.93(2.36) & 81.31(14.22) & 189.08(14.39) & 895.62(121.41)\\
	ICM/M & .08(.01) & 1.42(1.01) & 19.23(2.30) & 28.19(2.26) & 189.28(14.11) & 925.52(119.93)\\
	\hline
\end{tabular}
\end{table}

The results from Case 3 simulation were consistent with simulation studies in Case 1 and Case 2. ICM/M outperformed the othere two methods in terms of number of false positives although it has higher number of false negatives. ICM/M also had lowest misclassificationr rate in this case. For estimation errors, ICM/M and adaptive lasso also had similar ${\parallel}{\beta}{-}{\hat{\beta}}{\parallel}_1$ but adaptive lasso yielded smaller ${\parallel}{\beta}{-}{\hat{\beta}}{\parallel}_2^2$.

\begin{figure}
	\centering
	\includegraphics[scale=0.6]{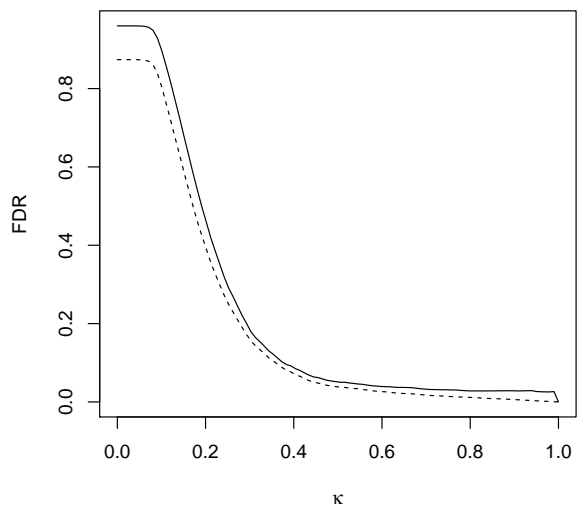}
	\vspace{-5 mm}
	\caption{$\kappa$ versus average true FDR (solid line) and estimated FDR (dash line) among 100 simulations in Case 3.} \label{figure:case3-FDRplot}
\end{figure}

As shown in Figure \ref{figure:case3-FDRplot}, FDR curve was well approximated in this case. In addition, the results in Table \ref{table:case3-result} were similar to the results in Table \ref{table:case3-kappa} when FDR was controlled at level 0.05.

\begin{table}[htbp]\footnotesize
	\centering
	\caption{Average $\kappa$, number of false positives (FP), number of false negatives (FN), and model sizes (MS) across 100 simulated data sets in Case 3 when controlling estimated FDR at level 0.05 (with standard deviation in parentheses) .} \label{table:case3-kappa}
	\begin{tabular}{ c c c c}
		\hline \hline
		$\kappa$ & FP & FN & No. of important variables \\
		\hline
		0.4604(0.0816) & 1.52(0.94) & 19.24(2.34) & 28.28(2.38)\\
		\hline
	\end{tabular}
\end{table}

To sum up the simulation section for binary logistic model, it is worth noticing that ICM/M with lasso fits as initial values improved significantly over lasso in terms of number of false positives. When correlation among predictors was very high, ICM/M might miss some important variables. However, it was interesting to see that ICM/M was able to reduce model size from lasso dramatically while increasing predictive ability at the same time. Furthermore, instead of relying on only regression coefficients from ICM/M, the local posterior probability along with the plot of estimated FDR can be used as a tool to select important variable.

\section{The ICM/M algorihm for Cox's proportional hazards model}
\subsection{Implementation details}
the classical survival analysis framework, the observed data consists of $(Y_i, \mathbf{X}_i, \delta_i)$ where $Y_i$ is the ovserved time for subject $i$, $\mathbf{X}_i$ is a $p$-dimensional vector of covariates, and $\delta_i$ is 1 if $Y_i$ is the actual survival time and 0 for right-censoring. The Cox's regression (\cite{Cox1972}) models survival times through hazard function
\begin{equation}
h(Y_i|\mathbf{X}_i) = h_0(Y_i)e^{\mathbf{X}_i\beta}
\end{equation}
where $h(Y_i|\mathbf{X}_i)$ is the hazard for subject $i$ and $h_0(Y_i)$ is the baseline hazard function (i.e., the hazard function when all covariates $\mathbf{X}_i$ are 0). The Cox's proportional hazard model assumes that covariates are time-independent and the baseline hazard $h_0(Y_i)$ is an unspecified function.\\

Let $t_1<t_2<...<t_m$ be the increasing list of distinct failure time and define cumulative baseline hazard at time $t_j$ for subject $i$ to be
\begin{equation}
H_0(Y_i) = \sum \limits_{j:t_j\le Y_i} \Delta H_0(t_j),
\end{equation}
where $\Delta H_0(t_j)$ is the increment of the cumulative baseline hazard at time $t_j$.\\

Following \cite{Johansen1983} and \cite{Nygard2008}, the regression coefficients can be found by maximizing the extended likelihood taking the form
\begin{equation}
L(\beta) = \prod \limits_{i=1}^n \left [ \{\Delta H_0(Y_i)e^{\mathbf{X}_i\beta}\}^{\delta_i} \exp \{-H_0(Y_i)e^{\mathbf{X}_i\beta}\}\right ].
\end{equation}
Given value of $\beta$, the increment of the cumulative baseline hazard that maximizes the likelihood is
\begin{equation}
\Delta \hat{H}_0(t_j) = \frac{\sum_{i:Y_i=t_j} \delta_i}{\sum_{i: Y_i\ge t_j e^{\mathbf{X}_i\beta}}}.
\end{equation}
Hence,
\begin{equation}
\hat{H}_0(Y_i) = \sum\limits_{j:t_j\le Y_i} \Delta \hat{H}_0(t_j).
\end{equation}
The explicit form of the pseudodata and pseudovariace can be computed based on current coefficients $\hat{\beta}$ as 
\begin{align}
Z_i & = \mathbf{X}_i \hat{\beta} + \frac{1}{\hat{H}_0(Y_i) e^{\mathbf{X}_i \beta}} \left ( \delta_i - \hat{H}_0(Y_i) e^{\mathbf{X}_i \hat{\beta}} \right )\\
\sigma_i^2 & = \frac{1}{\hat{H}_0(Y_i) e^{\mathbf{X}_i\hat{\beta}}}.
\end{align}
The approximated distribution of pseudodata $\mathbf{Z}$ is $N(\mathbf{X}\hat{\beta}, \Sigma)$ where $\Sigma = \text{diag}\{\sigma_i^2\}$. The ICM/M algorithm can now be used to cycle through parameters update and the active-set convergence is employed as stopping criterion.

\subsection{Simulation studies}
In this section, three cases in simulation were conducted to show the performance of the ICM/M algorithm for Cox's proportional hazards model. Here we compare the results with lasso and adaptive lasso for Cox's model in large $p$ small $n$ settings. In all three scenarios, we fixed $n=250$ and $p=1,000$. Survival times were simulated from a Cox model with the baseline hazard function of a Weibull distribution with a shape parameter $\nu=10$ and a scale parameter $\lambda=1$. The censoring times were generated randomly to achieve censoring rate of 50$\%$.

Case 4 simulation assumed independence among regression coefficients and ICM/M with independent prior in \eqref{eqn:indprior} was carried out. Case 5 and Case 6 assumed that regression coefficients followed some prior distribution representing the relationship among them. The ICM/M with the Ising prior was applied for Case 5 and Case 6 simulations. Here we report average number of false positives (FP), number of false negatives (FN), model sizes (MS), ${\parallel}{\beta}{-}{\hat{\beta}}{\parallel}_1 = \sum_{j-1}^p |\beta_j-\hat{\beta}_j|$, and ${\parallel}{\beta}{-}{\hat{\beta}}{\parallel}_2^2=\sum_{j=1}^p |\beta_j-\hat{\beta}_j|^2$ calculated among 100 simulated data sets for each case.

\subsubsection{Case 4: Independent prior}
The covariates were simulated the same way as in Case 1 with $n=250$ and $p=1,000$. Among 1,000 predictors, the failure times were determined by a linear combination of 20 non-zero coefficients: $\beta_1 = ...=\beta_{10}=5$ and $\beta_{101}=...=\beta_{110}=2$.  

\begin{table}[htbp]\footnotesize
	\centering
	\caption{Average number of false positives (FP), number of false negatives (FN), model sizes (MS), ${\parallel}{\beta}{-}{\hat{\beta}}{\parallel}_1$, and ${\parallel}{\beta}{-}{\hat{\beta}}{\parallel}_2^2$ across 100 simulated data sets in Case 4 (with standard deviation in parentheses) .} \label{table:case4-result}
	\begin{tabular}{ l | c c c c c}
		\hline \hline
		\multicolumn{6}{c}{$\rho = 0$} \\
		\hline 
		Method & FP & FN & MS &  ${\parallel}{\beta}{-}{\hat{\beta}}{\parallel}_1$ & ${\parallel}{\beta}{-}{\hat{\beta}}{\parallel}_2^2$\\
		\hline
		Lasso & 95.41(11.03) & 0.13(0.61) & 115.28(11.43) & 55.52(4.40) & 153.82(29.41)\\
		ALasso & 51.53(13.90) & 3.18(2.16) & 68.35(15.59) & 60.06(5.40) & 187.78(35.73)\\
		ICM/M & 0.03(0.17) & 0.81(2.11) & 19.22(2.07) & 50.98(4.74) & 153.15(27.18)\\
		\hline
		\multicolumn{6}{c}{$\rho = 0.3$} \\
		\hline
		Lasso & 105.16(6.33) & 0.00(0.00) & 125.16(6.33) & 46.22(4.49) & 100.40(24.17)\\
		ALasso & 56.14(12.87) & 1.32(1.48) & 74.99(14.09) & 48.81(8.65) & 125.81(48.00)\\
		ICM/M & 0.00(0.00) & 0.00(0.00) & 20.00(0.00) & 41.43(4.84) & 102.54(23.34)\\
		\hline
		\multicolumn{6}{c}{$\rho = 0.5$} \\
		\hline
		Lasso & 103.62(5.94) & 0.00(0.00) & 123.62(5.94) & 42.43(1.24) & 81.70(3.82)\\
		ALasso & 56.57(8.52) & 0.27(0.57) & 76.30(8.72) & 38.90(7.34) & 78.77(33.06)\\
		ICM/M & 0.01(0.10) & 0.00(0.00) & 20.01(0.10) & 37.62(0.86) & 83.67(3.88)\\
		\hline
		\multicolumn{6}{c}{$\rho = 0.7$} \\
		\hline
		Lasso & 98.78(6.09) & 0.00(0.00) & 118.78(6.09) & 41.43(4.27) & 79.30(19.71)\\
		ALasso & 41.42(8.45) & 0.06(0.28) & 61.36(8.43) & 35.84(3.89) & 68.53(16.36)\\
		ICM/M & 0.03(0.17) & 0.00(0.00) & 20.03(0.17) & 36.45(4.69) & 80.25(19.64)\\
		\hline
		\multicolumn{6}{c}{$\rho = 0.9$} \\
		\hline
		Lasso & 84.80(5.21) & 0.03(0.17) & 104.77(5.19) & 42.59(1.58) & 86.40(5.27)\\
		ALasso & 12.34(4.54) & 0.00(0.00) & 32.34(4.54) & 37.11(2.36) & 80.58(10.52)\\
		ICM/M & 0.17(0.40)  & 0.35(0.56) & 19.82(0.64) & 37.92(1.25) & 86.94(5.28)\\
		\hline
	\end{tabular}
\end{table}

Table \ref{table:case4-result} shows that ICM/M performed the best regarding the ability to select true important variables across all levels of correlation among covariates. The average numbers of both false positives and negatives were close to 0. Moreover, the average model size was close to the size of true model which was 20. Although lasso also had small numbers of false negatives, it turned out that lasso tended to select much larger number of non-zero coefficients which was more than 100 variables. This caused large numbers of false positives. Adaptive lasso selected moderate size of predictors but its numbers of false positives and negatives were relatively large comparing to those produced by ICM/M.

In a comparison of ${\parallel}{\beta}{-}{\hat{\beta}}{\parallel}_1$, ICM/M surpassed the other two methods for $\rho=0, 0.3,$ and $0.5$. For $\rho=0.7$ and $0.9$, adaptive lasso performed slightly better than ICM/M. When comparing estimation errors in terms of ${\parallel}{\beta}{-}{\hat{\beta}}{\parallel}_2^2$, adaptive lasso performed well when correlation among covariates was high while the performance of lasso and ICM/M were better than adaptive lasso for small value of $\rho$. Furthermore, we also observed that lasso and ICM/M yielded similar values of ${\parallel}{\beta}{-}{\hat{\beta}}{\parallel}_2^2$.

\begin{figure}[htp]
	\captionsetup[subfloat]{farskip=0pt,captionskip=0.1pt}
	\centering
	\subfloat[$\rho = 0$]{{\includegraphics[scale=0.65]{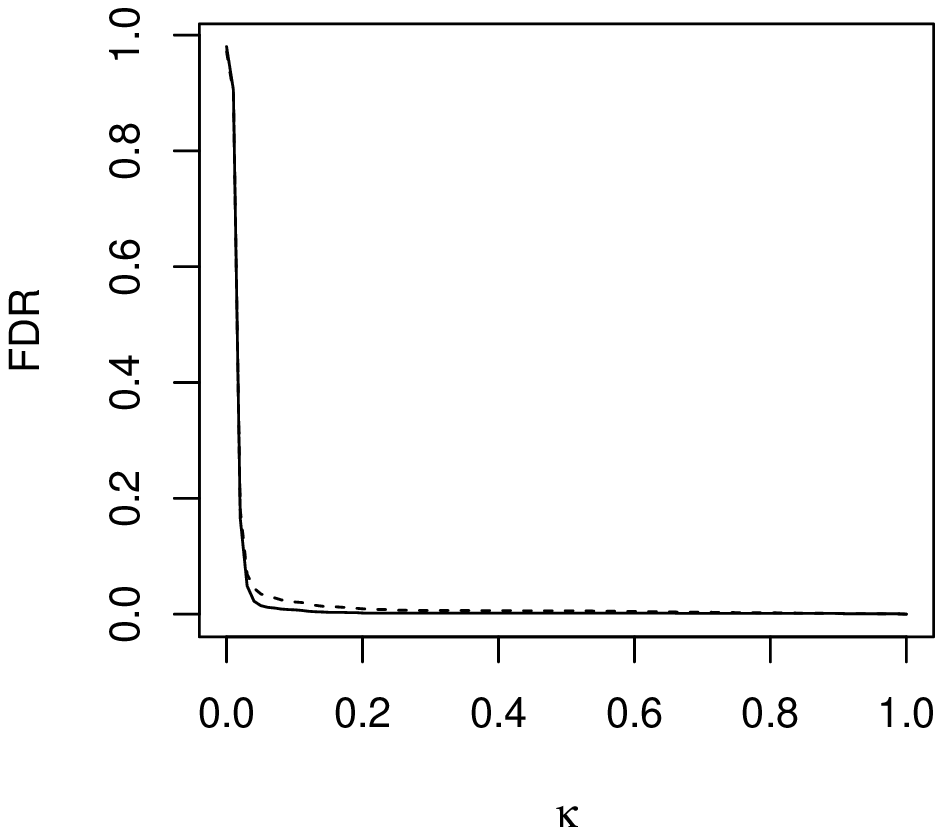} }} 
	\subfloat[$\rho=0.5$]{{\includegraphics[scale=0.65]{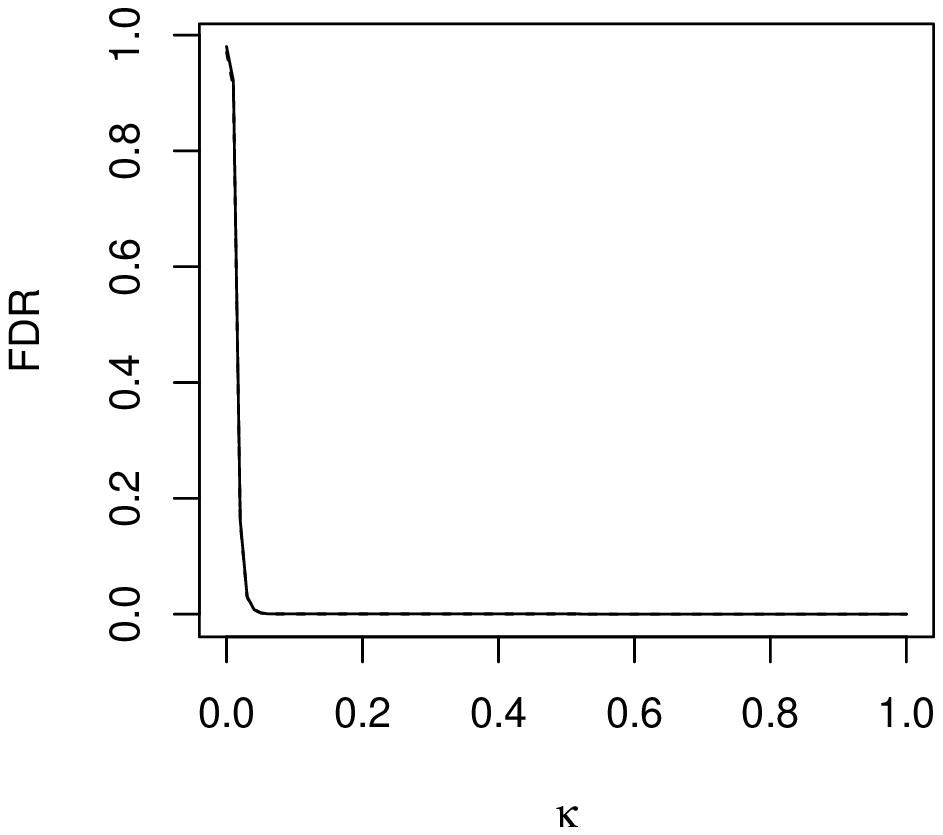} }}%
	\\[-0.8cm]
	\subfloat[$\rho=0.9$]{{\includegraphics[scale=0.65]{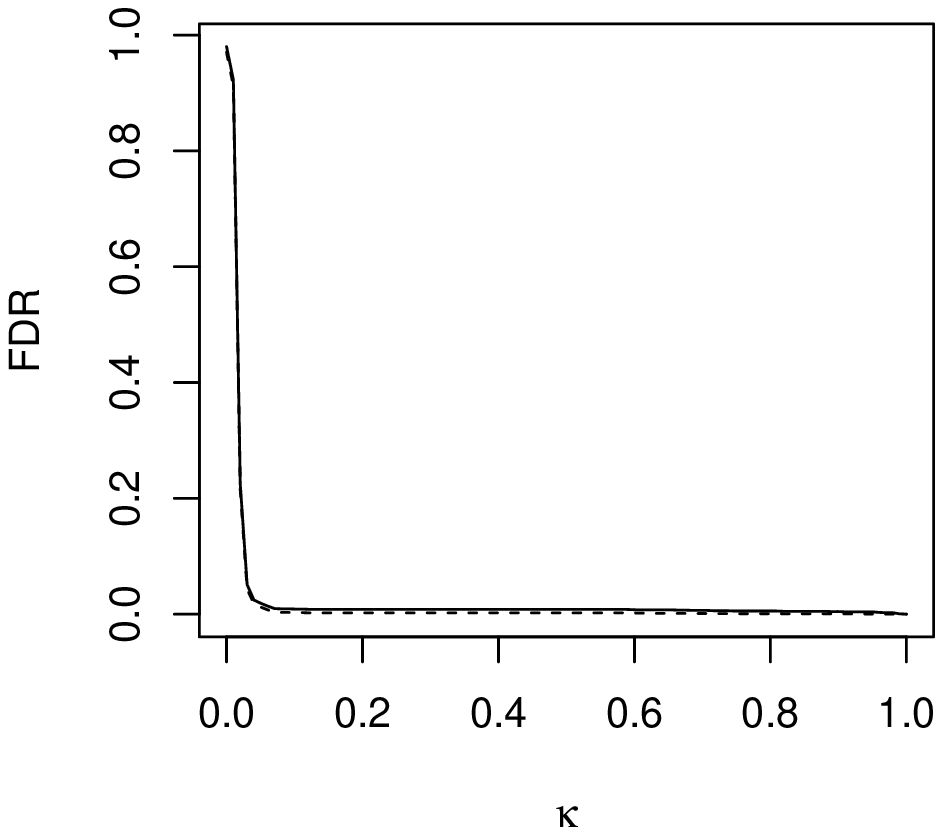} }}%
	\caption{$\kappa$ versus average true FDR (solid line) and estimated FDR (dash line) among 100 simulations in Case 4.} 
	\label{figure:case4-FDRplot}
\end{figure}

Figure \ref{figure:case4-FDRplot} shows that the estimated and true FDR curves were very close to each other. In addition, the results in Table \ref{table:case4-kappa} and Table \ref{table:case4-result} were consistent. The average number of false positives and negatives were all close to zero making the ICM/M was a competitive method for high-dimensional variable selection in Cox's model.

\begin{table}[htbp]\footnotesize
	\centering
	\caption{Average $\kappa$, number of false positives (FP), number of false negatives (FN), and model sizes (MS) across 100 simulated data sets in Case 4 when controlling estimated FDR at level 0.05 (with standard deviation in parentheses) .} \label{table:case4-kappa}
	\begin{tabular}{ c | c c c c}
		\hline \hline
		$\rho$ & $\kappa$ & FP & FN & No. of important variables \\
		\hline
		0 & 0.0681(0.1236) & 0.30(0.46) & 0.81(2.21) & 19.49(2.32)\\
		0.3 & 0.0283(0.0061) & 0.46(0.50) & 0.00(0.00) & 20.46(0.50)\\
		0.5 & 0.0302(0.0070) & 0.42(0.50) & 0.00(0.00) & 20.42(0.50)\\
		0.7 & 0.0303(0.0070) & 0.46(0.50) & 0.00(0.00) & 20.46(0.50)\\
		0.9 & 0.0330(0.0123) & 0.50(0.59) & 0.35(0.56) & 20.15(0.86)\\
		\hline
	\end{tabular}
\end{table}

\subsubsection{Case 5: Linear chain prior}
Covariates with $n=250$ and $p=1,000$ were generated the same way as in Case 2. The location of non-zero coefficients follows a Markov linear chain with the same transition matrix as in Case 2. The effect sizes of those non-zero coefficients were drawn from Uniform[0.5,5]. 

\begin{table}[htbp]\footnotesize
	\centering
	\caption{Average number of false positives (FP), number of false negatives (FN), model sizes (MS), ${\parallel}{\beta}{-}{\hat{\beta}}{\parallel}_1$, and ${\parallel}{\beta}{-}{\hat{\beta}}{\parallel}_2^2$ across 100 simulated data sets in Case 5 (with standard deviation in parentheses) .} \label{table:case5-result}
	\begin{tabular}{ l | c c c c c}
		\hline \hline
		\multicolumn{6}{c}{$\rho = 0$} \\
		\hline 
		Method & FP & FN & MS &  ${\parallel}{\beta}{-}{\hat{\beta}}{\parallel}_1$ & ${\parallel}{\beta}{-}{\hat{\beta}}{\parallel}_2^2$\\
		\hline
		Lasso & 94.30(7.82) & 0.29(0.50) & 113.01(7.81) & 45.89(2.16) & 104.40(10.45)\\
		ALasso  & 59.09(12.01) & 2.00(1.67) & 76.09(13.17) & 45.98(5.71) & 108.68(30.44)\\
		ICM/M & 0.01(0.10) & 0.63(0.66) & 18.38(0.65) & 40.58(2.09) & 99.19(9.35)\\
		\hline
		\multicolumn{6}{c}{$\rho = 0.3$} \\
		\hline
		Lasso & 98.53(8.76) & 0.11(0.35) & 117.42(8.79) & 41.96(4.09) & 84.62(20.28)\\
		ALasso & 57.18(11.67) & 1.26(1.33) & 74.92(12.67) & 42.81(6.11) & 94.91(30.11)\\
		ICM/M & 0.07(0.29) & 0.39(0.62) & 18.68(0.65) & 36.45(4.49) & 80.97(19.16)\\
		\hline
		\multicolumn{6}{c}{$\rho = 0.5$} \\
		\hline
		Lasso & 97.52(6.03) & 0.10(0.30) & 116.42(6.02) & 41.15(2.01) & 80.21(9.13)\\
		ALasso  & 56.94(11.35) & 0.90(1.13) & 75.04(12.15) & 40.29(6.72) & 84.11(31.80)\\
		ICM/M & 0.10(0.33) & 0.44(0.57) & 18.66(0.67) & 35.85(2.08) & 77.18(8.74)\\
		\hline
		\multicolumn{6}{c}{$\rho = 0.7$} \\
		\hline
		Lasso & 95.19(7.20) & 0.15(0.36) & 114.04(7.17) & 39.59(3.76) & 73.77(17.78)\\
		ALasso  & 52.45(10.02) & 1.00(1.13) & 70.45(10.64) & 39.94(6.29) & 82.17(28.44)\\
		ICM/M & 0.50(0.73) & 0.55(0.61) & 18.95(0.86) & 34.36(4.19) & 71.43(17.16)\\
		\hline
		\multicolumn{6}{c}{$\rho = 0.9$} \\
		\hline
		Lasso & 83.65(6.03) & 0.41(0.59) & 102.24(5.99) & 39.70(1.92) & 71.06(6.47)\\
		ALasso & 47.19(7.09) & 0.99(0.97) & 65.20(7.26) & 35.25(5.30) & 58.83(19.04)\\
		ICM/M & 1.97(1.62) & 0.88(0.82) & 20.09(1.80) & 34.41(1.67) & 68.98(6.23)\\
		\hline
	\end{tabular}
\end{table}

Ability to select variables correctly were assessed in 3 criteria: number of false positives, false negatives, and model size. As you can see in Table 5, ICM/M was the best method to select correct variables here due to small numbers of false positives and false negatives across all value of $\rho$.  Average model sizes were also close to the true model size containing 19 non-zero coefficients. We also noticed that ICM/M gave slightly higher number of false positives when $\rho$ increased. In contrast, lasso and adaptive lasso yielded lower number of false positives when $\rho$ increased. Although lasso and adaptive lasso performed reasonable well with regard to number of false negatives, both methods always selected large number of unimportant variables resulting in high false positive rates. Besides the variable selection ability, ICM/M overall outperformed the other two methods regarding to the two criteria of estimation errors including ${\parallel}{\beta}{-}{\hat{\beta}}{\parallel}_1$ and ${\parallel}{\beta}{-}{\hat{\beta}}{\parallel}_2^2$. Numbers of important variables in Table \ref{table:case5-kappa} when controlling FDR at level 0.05 were slightly higher than the model size in Table \ref{table:case5-result} but still very close to the true number of important variables.

\begin{figure}[htp]
	\captionsetup[subfloat]{farskip=0pt,captionskip=0.1pt}
	\centering
	\subfloat[$\rho = 0$]{{\includegraphics[scale=0.65]{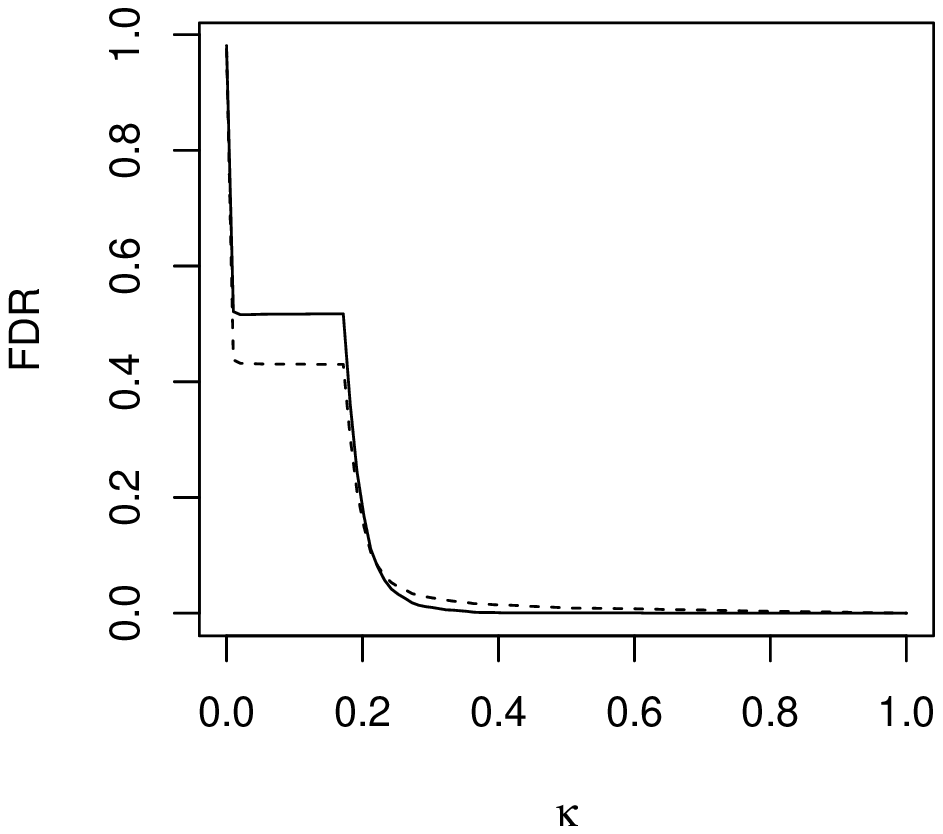} }} 
	\subfloat[$\rho=0.5$]{{\includegraphics[scale=0.65]{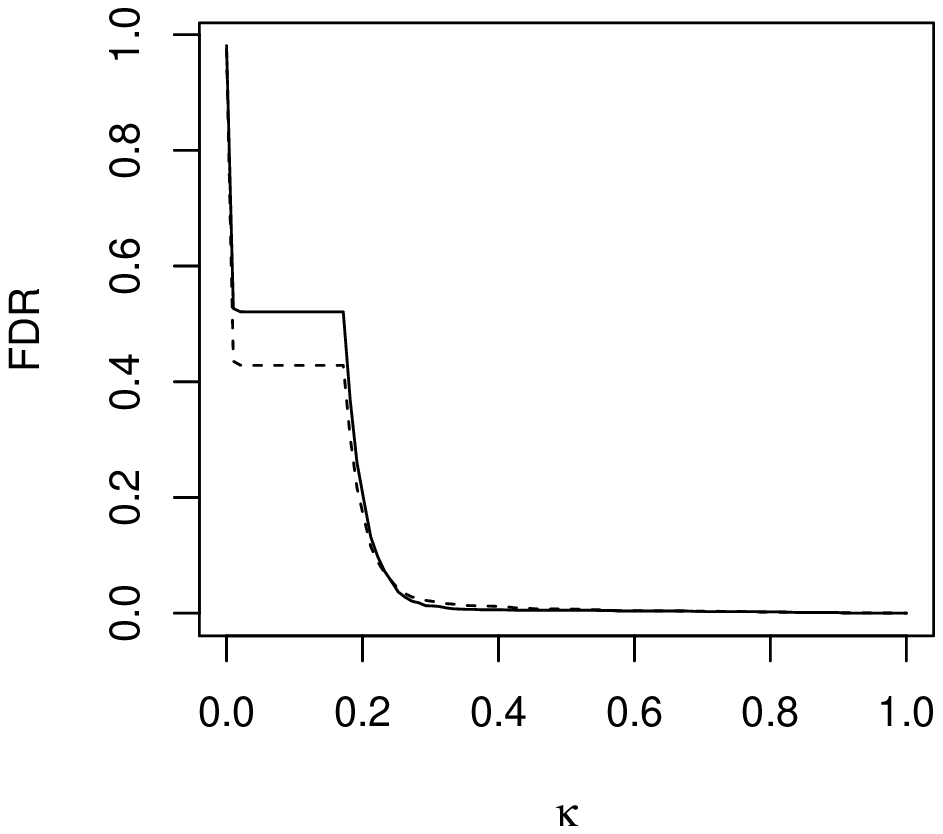} }}%
	\\[-0.8cm]
	\subfloat[$\rho=0.9$]{{\includegraphics[scale=0.65]{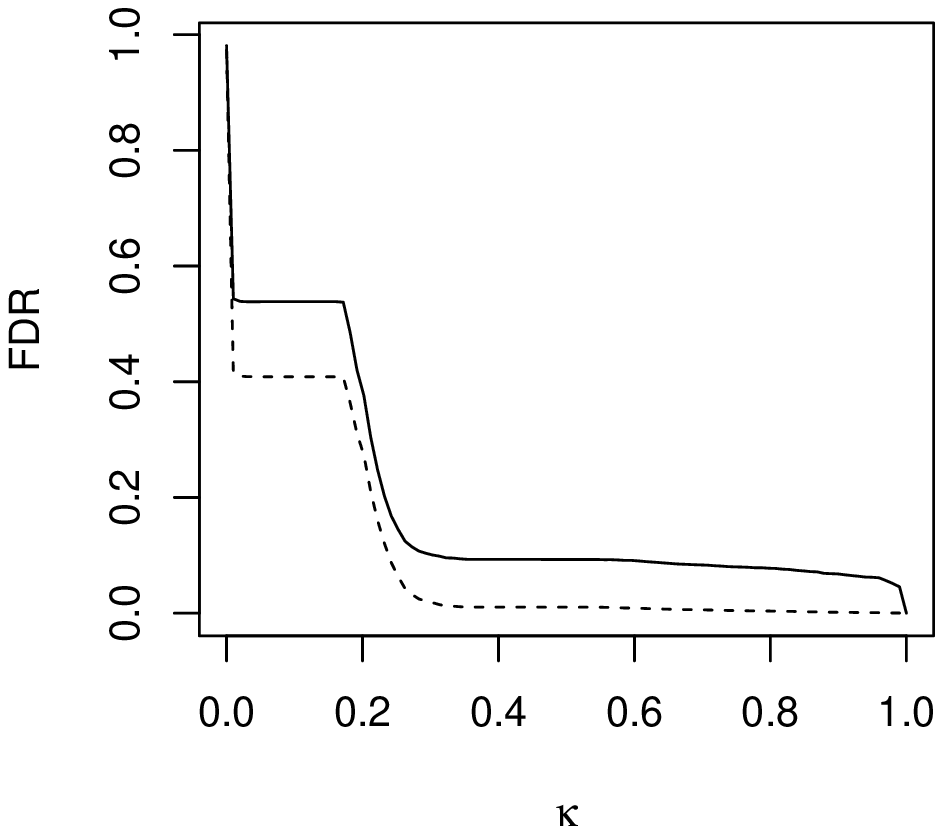} }}%
	\caption{$\kappa$ versus average true FDR (solid line) and estimated FDR (dash line) among 100 simulations in Case 5.} 
	\label{figure:case5-FDRplot}
\end{figure}

\begin{table}[htbp]\footnotesize
	\centering
	\caption{Average $\kappa$, number of false positives (FP), number of false negatives (FN), and model sizes (MS) across 100 simulated data sets in Case 5 when controlling estimated FDR at level 0.05 (with standard deviation in parentheses) .} \label{table:case5-kappa}
	\begin{tabular}{ c | c c c c}
		\hline \hline
		$\rho$ & $\kappa$ & FP & FN & No. of important variables \\
		\hline
		0 & 0.2554(0.0408) & 0.34(0.48) & 0.26(0.56) & 19.08(0.66)\\
		0.3 & 0.2506(0.0474) & 0.59(0.55) & 0.16(0.47) & 19.43(0.74)\\
		0.5 & 0.2542(0.0498) & 0.48(0.56) & 0.10(0.39) & 19.38(0.68)\\
		0.7 & 0.2572(0.0465) & 0.87(0.82) & 0.28(0.51) & 19.59(0.83)\\
		0.9 & 0.2509(0.0383) & 2.37(1.71) & 0.73(0.78) & 20.64(1.87)\\
		\hline
	\end{tabular}
\end{table}

\subsubsection{Case 6: Pathway structure prior}
For Case 6, gene expression data within an assumed network were simulated. The network consisted of 10 disjoint pathways. Each of which contained 100 genes resulting in $p=1,000$ in total. Ten regulated genes were assumed in each pathway. The gene expression values were generated from a standard normal distribution. For those regulated genes in the same pathway, the expression values were generated from normal distribution with a correlation of $\rho=0.7$ among those 10 regulated genes. Eighteen regulated genes from three pathways were chosen to have non-zero coefficients that were drawn from Uniform[0.5, 5].

\begin{table}[htbp]\footnotesize
	\centering
	\caption{Average number of false positives (FP), number of false negatives (FN), model sizes (MS), ${\parallel}{\beta}{-}{\hat{\beta}}{\parallel}_1$, and ${\parallel}{\beta}{-}{\hat{\beta}}{\parallel}_2^2$ across 100 simulated data sets in Case 6 (with standard deviation in parentheses) .} \label{table:case6-result}
	\begin{tabular}{ l | c c c c c}
		\hline \hline
		Method & FP & FN & MS &  ${\parallel}{\beta}{-}{\hat{\beta}}{\parallel}_1$ & ${\parallel}{\beta}{-}{\hat{\beta}}{\parallel}_2^2$\\
		\hline
		Lasso & 86.50(6.34) & 0.01(0.10) & 104.49(6.33) & 29.95(2.34) & 45.66(9.21)\\
		ALasso  & 31.03(7.71) & 0.00(0.00) & 49.03(7.71) & 20.39(2.70) & 23.72(7.17)\\
		ICM/M & 0.45(0.64) & 0.59(0.74) & 17.86(0.89) & 26.00(2.67) & 44.25(8.96)\\
		\hline
	\end{tabular}
\end{table}

As you can see in Table \ref{table:case6-result}, ICM/M outperformed the other two methods in terms of the ability to select true relevant genes. Both average numbers of false positives and negatives were close to 0 with the average model size being 17.86 which is close to 18. Lasso and adaptive lasso selected much more variables into the model resulting in large number of false positives. We also observed that adaptive lasso had smallest avarage values for ${\parallel}{\beta}{-}{\hat{\beta}}{\parallel}_1$ and ${\parallel}{\beta}{-}{\hat{\beta}}{\parallel}_2^2$. 

\begin{figure}
	\centering
	\includegraphics[scale=0.6]{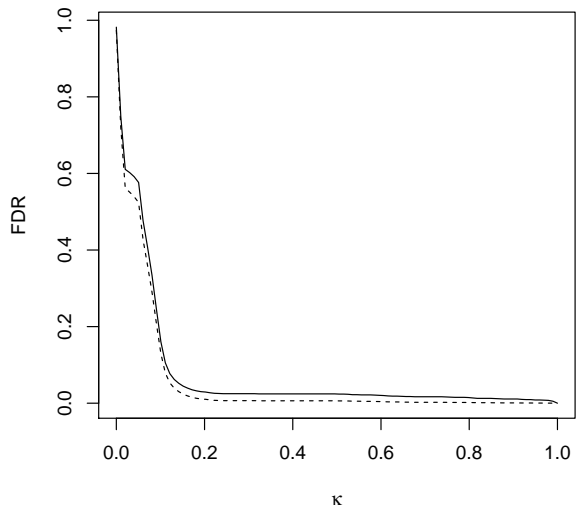}
	\vspace{-5 mm}
	\caption{$\kappa$ versus average true FDR (solid line) and estimated FDR (dash line) among 100 simulations in Case 6.} \label{figure:case6-FDRplot}
\end{figure}

Similar to other simulation studies for Cox's model, the true FDR curve was well approximated by the estimated FDR curve. Moreover, the results based on the regression coefficients from ICM/M method in Table \ref{table:case6-result} were similar to variable selection procedure at FDR level = 0.05 in Table \ref{table:case6-kappa}.

\begin{table}[htbp]\footnotesize
	\centering
	\caption{Average $\kappa$, number of false positives (FP), number of false negatives (FN), and model sizes (MS) across 100 simulated data sets in Case 6 when controlling estimated FDR at level 0.05 (with standard deviation in parentheses) .} \label{table:case6-kappa}
	\begin{tabular}{ c c c c}
		\hline \hline
		$\kappa$ & FP & FN & No. of important variables \\
		\hline
		0.1259(0.0311) & 0.88(0.76) & 0.59(0.74) & 18.29(0.95)\\
		\hline
	\end{tabular}
\end{table}

To conclude the simulation results for Cox's proportional hazards model, ICM/M gave the best variable selection performance. Lasso and adaptive lasso tended to select much larger number of variables into the model yielding large value of false positives though they might have lower number of false negatives than ICM/M. This suggested the substantial improvement of ICM/M with lasso coefficients as initial values over lasso. In addition, we also noticed that the results based on ICM/M coefficients were comparable to variable selection based on the local posterior probability when FDR is being controlled at level 0.05.

\section{Illustration: Parkinson's disease genome-wide association study}
As an application in genomics, a GWAS was carried out to identify a set of susceptible genetic markers to Parkinson's disease (PD). This PD dataset is publicly available (dbGaP study accession number: phs000089.v3.p2). The dataset consisted of 940 PD patients and 801 controls resulting in 1,741 samples in total. All individuals were Americans with European ancestry. The genotyping assays were derived from Illumina Infinium HumanHap300 and HumanHap500 SNP chips. There were 310,860 SNPs in common between these two arrays which we focused in this analysis. The following GWAS preprocessing criteria were applied to the data: (i) missingness per individual $< 10\%$, (ii) missingness per marker $< 10\%$, (iii) minor allele frequency (MAF) $\ge$ 5$\%$, and (iv) Hardy-Weinberg test at significant level = 0.001. In addition, K-nearest neighbor (KNN) method was employed to impute missing genotype.
Parkinson metabolic pathway and other six pathways related to PD were acquired from the KEGG database. There were 26,101 SNPs representing 341 genes in the seven pathways. To make the computation more manageable, the univariate tests were applied and only 2,664 SNPs having p-values surpassed the significant threshold at 0.1.

Assume known pathway information among SNPs from the KEGG database, ICM/M with Ising prior was applied to fit a high-dimensional binary logistic regression. Again, lasso fits were used as initial values of coefficients for ICM/M algorithm. ICM/M was then compared with lasso and adaptive lasso. 

From Table \ref{table:PD-result}, Lasso performed relatively poor due to large misclassification rate and model size. Adaptive lasso had the smallest misclassification rate. Similar to results in simulation studies, ICM/M tended to select smaller number of variables into the model comparing to lasso and adaptive lasso. There was also an improvement of ICM/M in terms of misclassification rate over lasso.

\begin{table}[htbp]\footnotesize
	\centering
	\caption{Misclassification rate and number of identified SNPs for PD data analysis}
	\label{table:PD-result}
	\begin{tabular}{ l | c c }
		\hline \hline
		Method & Misclassification rate & No. of identified SNPs \\
		\hline
		Lasso & 0.1344 & 344 \\
		ALasso & 0.0677 & 266 \\
		ICM/M & 0.0781  & 170 \\
		\hline
	\end{tabular}
\end{table}

\begin{figure}
	\centering
	\includegraphics[scale=0.65]{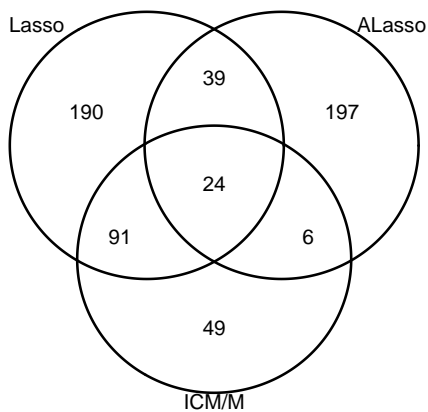}
	\caption{Number of SNPs identified by lasso, adaptive lasso, and ICM/M for PD data analysis.} \label{figure:PD-overlappingSNPs}
\end{figure}

Figure \ref{figure:PD-overlappingSNPs} demonstrates the number of SNPs identified by the three methods. There were 24 SNPs identified by all three methods. The details of these overlapping SNPs are shown in Table \ref{table:PD-overlappingSNPs}. We noticed that the signs of $\hat{\beta}$ for each SNPs were the same for three methods. Overall, ICM/M tended to produce larger effect sizes for regression coefficients than the other two methods. This might due to the thresholding of ICM/M to screen out the small regression coefficients.

\begin{table}[htbp]\footnotesize
	\centering
	\caption{SNPs identified by all three methods: lasso, adaptive lasso, and ICM/M in PD data analysis.} 
	\label{table:PD-overlappingSNPs}
	\begin{tabular}{ c c c | c c c | c }
		\hline \hline
		\multirow{2}{*}{Chr.} & \multirow{2}{*}{SNP} & \multirow{2}{*}{Physical Location} & \multicolumn{3}{|c|}{$\hat{\beta}$} & \multicolumn{1}{|c}{$\zeta$}\\
		& & & Lasso & ALasso & ICM/M & ICM/M \\
		\hline
		1 & rs4240910 & 9690399 & 0.0832 & 0.1455 & 0.2449 & 0.9988\\
		3 & rs769801 & 3730854 & -0.0074 & -0.1529 & -0.1465 & 0.8676\\ 	
		3 & rs162216 & 7816361 & -0.0558 & -0.0712 & -0.2519 & 0.9982\\
		3 & rs6788714 & 11705263 & -0.0305 & -0.0772 & -0.2032 & 0.9795\\
		3 & rs2133060 & 11782845 & 0.0390 & 0.0263 & 0.2226 & 0.9919\\
		3 & rs2731938 & 21354604 & -0.0147 & -0.0655 & -0.2121 & 0.9888\\
		3 & rs3860583 & 21629028 & -0.0014 & -0.0441 & -0.1703 & 0.9313\\
		3 & rs6792725 & 24478792 & 0.0246 & 0.0879 & 0.1632 & 0.9155\\
		3 & rs12152294 & 25174179 & 0.0357 & 0.0317 & 0.1766 & 0.9436\\
		3 & rs1982639 & 27900063 & 0.0134 & 0.0763 & 0.1842 & 0.9603\\
		4 & rs2736990 & 89757390 & -0.0288 & -0.1074 & -0.1416 & 0.9010\\
		6 & rs9378352 & 3010915 & -0.0299 & -0.0742 & -0.1552 & 0.8918\\
		6 & rs9350088 & 10095223 & -0.0138 & -0.1225 & -0.1218 & 0.7934\\
		6 & rs4715167 & 13220090 & -0.0055 & -0.1821 & -0.1035 & 0.7294\\
		6 & rs554400 & 23997929 & 0.0073 & 0.0311 & 0.1716 & 0.9238\\
		7 & rs999228 & 138716113 & -0.0124 & -0.0117 & -0.1382 & 0.8465\\
		17 & rs1369112 & 33604573 & 0.0248 & 0.0701 & 0.1964 & 0.9776\\
		17 & rs146891 & 36765872 & 0.01666 & 0.0269 & 0.1549 & 0.9112\\
		17 & rs2289672 & 44854876 & -0.0333 & -0.0051 & -0.2352 & 0.9975\\
		17 & rs919089 & 49970057 & -0.0196 & -0.0854 & -0.1975 & 0.9816\\
		17 & rs4968723 & 64339579 & -0.0667 & -0.0091 & -0.1806 & 0.9584\\
		17 & rs1991401 & 64506317 & 0.0370 & 0.1018 & 0.2146 & 0.9910\\
		17 & rs4447484 & 68725298 & -0.0001 & -0.0643 & -0.1953 & 0.9771\\
		17 & rs1469587 & 70735987 & 0.0380 & 0.0234 & 0.1449 & 0.8872\\
		\hline
	\end{tabular}
\end{table}

For ICM/M, FDR obtained from the local posterior probability was estimated. When controlling FDR at level 0.05, $\kappa$ was chosen to be 0.86 yielding 114 important variables. If FDR level was increased to 0.10, the corresponding $\kappa = 0.64$ yielding 157 important variables. 

Due to large number of identified SNPs, we further investigated our analysis. For comparison purposes, we only focused on the results of ICM/M based on the fitted model for the rest of this section although one might consider using FDR and the local posterior probability to select important variables instead.

Since the correlation among SNPs resided in the same genetic region tended to be high, it was possible that different variable selection methods select different sets of SNPs representing the same gene. We therefore obtained genetic markers by grouping SNPs affiliated to the same gene. The genetic marker for gene $g$ was defined as $\tilde{\mathbf{X}}_g = \mathbf{X}_g \hat{\beta}_g$ where $\mathbf{X}_g$ was a design matrix consisting of all SNPs resided in gene $g$ and $\hat{\beta}_g$ was a vector of corresponding estimated coefficients obtained from variable selection method when considered all SNPs in original data as predictors. With the construction of genetic markers, it was easy to check which genes revealed genetic risk to PD. For each method, the logistic regression model with disease status as a response and genetic markers as covariates was fitted to obtain the p-values. The numbers of identified SNPs within the genetic regions are reported here. It was possible that some SNPs were affiliated to more than one gene. Table \ref{table:PD-identifiedgenes} provides the results for genes identified by ICM/M (Full table can be found in Supplementary Table A).

\begin{table}[htbp]\footnotesize
	\centering
	\caption{Genes identified by ICM/M in PD data analysis (Bolded genes were identified by all methods).} \label{table:PD-identifiedgenes}
\begin{tabular}{l | c c c | c c c }
	\hline
	\hline
	\multirow{2}{*}{Gene (Location)} & \multicolumn{3}{|c|}{Number of identified SNPs} & \multicolumn{3}{|c}{p-value} \\
	&  Lasso & ALasso & ICM/M & Lasso & ALasso & ICM/M \\
	\hline
	ATP5F1 (1p13.2) & 1 & 0 & 1 & 0.0088 & 1 &  1.01e-6 \\
	NDUFS2 (1q23) & 0 & 0 & 1 & 1 & 1 & 0.0010 \\
	\textbf{PIK3CD} (1p36.2) & 1 & 1 & 1 & 5.93e-8 & 0.0037 & 1.14e-8 \\
	BIRC6 (2p22.3) & 2 & 0 & 1 & 5.23e-8 & 1 & 2.13e-9 \\
	\textbf{TRIP12} (2q36.3) & 1 & 1 & 1 & 0.0352 & 0.0638 & 4.51e-5 \\
	\textbf{NDUFB4} (3q13.33) & 159 & 141 & 62 & 1.16e-70 & 3.2e-58 & 8.69e-66 \\	\textbf{HGD} (3q13.33) & 159 & 141 & 62 & 1.16e-70 & 3.2e-58 & 8.69e-66 \\
	PRKAR2A (3p21.3-p21.2) & 0 & 0 & 1 & 1 & 1 & 0.9480 \\
	IL1RAP (3q28) & 2 & 0 &1 & 0.0021 & 1 & 3.17e-5 \\
	UCHL1 (4p14) & 0 & 1 & 1 & 1 & 0.0317 & 0.0146 \\
	\textbf{SNCA} (4q21) & 1 & 1 & 1 & 0.0043 & 0.0029 & 0.0020 \\
	\textbf{ATP6V1G2} (6p21.3) & 48 & 38 & 17 & 2.41e-28 & 5.99e-24 & 1.26e-38 \\
	\textbf{TNF} (6p21.3) & 48 & 38 & 17 & 2.41e-28 & 5.99e-24 & 1.26e-38 \\
	\textbf{PARK2} (6q25.2-q27) & 5 & 5 & 3 & 3.29e-8 & 2.42e-5 & 1.34e-17 \\
	PRKAR2B (7q22) & 1 & 0 & 1 & 6.00e-4 & 1 & 7.01e-07 \\
	\textbf{ATP6V0A4} (7q34) & 2 & 1 & 1 & 0.0098 & 0.0934 & 0.0002 \\
	TNFRSF10B (8p22-p21) & 1 & 0 & 1 & 0.0205 & 1 & 0.0183 \\
	\textbf{CUL2} (10p11.21) & 1 & 1 & 1 & 0.0339 & 0.0034 & 0.0005 \\
	FAS (10q24.1) & 0 & 0 & 1 & 1 & 1 & 0.0044 \\
	\textbf{PRMT3} (11p15.1) & 1 & 2 & 1 & 3.97e-5 & 0.0001 & 2.58e-7 \\
	\textbf{MAPT} (17q21.1) & 107 & 51 & 72 & 7.11e-52 & 2.78e-28 & 6.17e-66 \\
	\textbf{ANAPC11} (17q25.3) & 107 & 51 & 72 & 7.11e-52 & 2.78e-28 & 6.17e-66 \\
	NDUFA11 (19p13.3) & 1 & 0 & 1 & 0.0118 & 1 & 8.15e-5 \\
	ATP5O (21q22.11) & 0 & 0 & 1 & 1 & 1 & 7.98e-9 \\
	\hline
\end{tabular}
\end{table}

There were 28, 35, and 24 genes identified by lasso, adaptive lasso, and ICM/M respectively (See Supplementary Table A). Among 13 genes detected for all three methods, 4 of them including MAPT, PARK2, PIK3CD, and SNCA had been well studied and known as PD susceptible genes in a number of literatures (see e.g., \cite{Pankratz2009}, \cite{Simon-Sanchez2009}, \cite{Klien2012}, \cite{Polito2016}, and \cite{Lill2016}). There were another 6 genes that had been reported in Harmonizome database (see \cite{Rouillard2016}) to have association to PD based on GWAS and other genetic association datasets from the GWASdb SNP-Disease Associations dataset. These 6 genes included BIRC6, COX7B2, PRKAR2B, TNPO3, and TRIP12. Among these 6 genes, lasso, adaptive lasso, and ICM/M can identify 5, 4, and 4 genes respectively. In addition, UCHL1 had also been reported to be involved with PD risk in recent literatures (\cite{Andersson2011} and \cite{Hernandez2016}). Here, only ICM/M and adaptive lasso are able to detect UCHL1 gene.

We observed that there was a relationship between p-values and number of identified SNPs resided in the genetic region. Indeed, the genes having larger number of identified SNPs tended to have smaller p-values. This phenomenon was not unusual due to the fact that genes with larger number of identified SNPs were accounted for more variations to the response. Interestingly, we noticed that the p-value for ICM/M method tended to be smaller than the other two methods despite the number of identified SNPs resided in a gene. 
		
\section{Illustration: Lung Adenocarcinoma Microarray Analysis}
We illustrate our proposed methodology for Cox's model to the lung cancer data from \cite{Beer2002}. The microarray data consisting of $p=7,129$ genes were from $n=86$ patients with primary lung adenocarcinomas. The survival time of patients in early-stage lung adenocarinomas were recorded and the censoring rate in this data set is $27.9\%$. 

Three pathways related to lung adenocarcinoma were obtained from KEGG database. Genes in the data set were then mapped to genes in these three pathways to generate an undirectedg graph. This resulted in a gene network consisting of 55 nodes and 109 edges. The rest of the genes were either isolated or not in one of these three pathways. Using lasso coefficients as initial values, the ICM/M algorithm with pathway structure prior was applied to the data set. For comparison purposes, the results from lasso and adaptive lasso were also reported.

\begin{figure}
	\centering
	\includegraphics[scale=0.65]{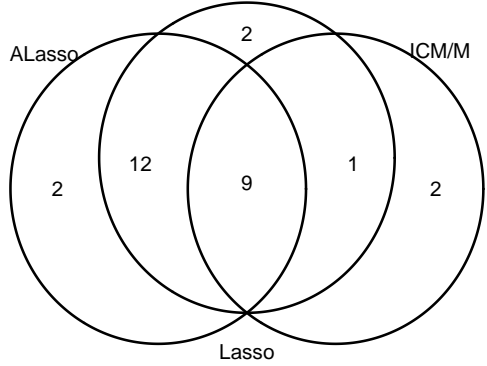}
	\vspace{-1cm}
	\caption{Number of genes identified by lasso, adaptive lasso, and ICM/M.}
	 \label{figure:lung-overlappinggenes}
\end{figure}

Lasso , adaptive lasso, and ICM/M selected 24, 23, and 12 genes respectively. Figure \ref{figure:lung-overlappinggenes} shows the number of overlapping genes that were identified by the three methods. The numbers of selected genes by lasso and adaptive lasso were similar and there were 21 overlapping genes between these two methods. On the other hands, ICM/M selected less genes than the other two methods. There were 9 genes identified by all three methods. Among these 9 genes, 4 of them including BAG1, FXYD3, HPIP, and SLC2A1 were among the top 100 genes related to survival of patients reported in Beer et. al. (2002) (see Table \ref{table:lung-result}). Furthermore, we found that another 3 genes were reported to have some relation with lung cancer in literatures. This included INSL4 (\cite{Ludovini2016}), PRKACB (\cite{Chen2013}), and SPRR1B(\cite{Ludovini2016}).

\begin{table}[htbp]\footnotesize
	\centering
	\caption{Genes identified by all three methods: lasso, adaptive lasso, and ICM/M in  lung adenocarcinoma microarray analysis (Bolded genes were among top 100 genes reported in \cite{Beer2002}).}
	\label{table:lung-result}
	\begin{tabular}{ l c | c c c | c }
		\hline \hline
		 \multirow{2}{*}{Gene} & \multirow{2}{*}{Location} & \multicolumn{3}{|c|}{$\hat{\beta}$} & \multicolumn{1}{|c}{$\zeta$}\\
		 & & Lasso & ALasso & ICM/M & ICM/M \\
		\hline
		\textbf{BAG1} & 9p13.3 & 0.1179 & 0.2274 & 0.3684 & 0.8937 \\
		\textbf{FXYD3} & 19q13.12 & 0.1088 & 0.0536 & 0.1658 & 0.6469 \\
		\textbf{HPIP} & 1q21.3 & -0.1287 & -0.2308 & -0.3069 & 0.7962 \\
		INSL4 & 9p24.1 & 0.2159 & 0.3268 & 0.3243 & 0.9383 \\
		PRKACB & 1p31.1 & -0.2727 & -0.2873 & -1.2073 & 1.0000\\
		SLAM & 1q23.3 & -0.3141 & -0.1177 & -1.0650 & 1.0000\\
		\textbf{SLC2A1} & 1p34.2 & 0.3185 & 0.3893 & 0.6393 & 0.9993\\
		SPRR1B & 1q21.3 & 0.0793 & 0.1338 & 0.0959 & 0.6169\\
		Y00477 & 19p13.3 & -0.3117 & -0.3494 & -0.7525 & 0.9998\\
		\hline
	\end{tabular}
\end{table}

In addition to 9 genes in Table \ref{table:lung-result}, ICM/M also identified 3 more genes including NME2, STMN1, and ZFP36 (see Table \ref{table:lung-threegenesbyICMM}). In fact, NME2, identified by both Lasso and ICM/M, was among the top 100 genes related to survival of patients reported in Beer et. al. (2002). For STMN1 and ZFP36, only ICM/M was able to identify these 2 genes. Interestingly, both genes were related to lung cancer in recent literatures: STMN1 (\cite{Nie2015}) and ZFP36 (\cite{Shao2017}). 

\begin{table}[htbp]\footnotesize
	\centering
	\caption{Three more genes identified by ICM/M in  lung adenocarcinoma microarray analysis (Bolded genes were among top 100 genes reported in \cite{Beer2002}).}
	\label{table:lung-threegenesbyICMM}
	\begin{tabular}{ l c | c c }
		\hline \hline
		Gene & Location & $\hat{\beta}$ & $\zeta$\\
		\hline
		\textbf{NME2} & 17q21.33 & 0.3468 & 0.8008 \\
		STMN1 & 1p36.11 & -0.0385 & 0.5655 \\
		ZFP36 & 19q13.2 & 0.1302 & 0.6125\\
		\hline
	\end{tabular}
\end{table}

For ICM/M method, estimated FDR were calculated and $\kappa$ was chosen as 0.81 which corresponded to FDR level 0.05. This yielded 6 genes having local posterior probabilities surpass $\kappa=0.81$. If one would like to increase the number of selected genes, FDR level can be increased to any desired level. For example, by controlling FDR level at 0.10, the $\kappa$ was chosen to be 0.66 yielding 8 important genes. To select all 12 genes, the corresponding FDR level was 0.15 in this case. 

\section{Discussion}
\cite{Pungpapong2015} proposed an empirical Bayes methods in selecting massive variables for linear model. A apike-and-slab prior is employed to introduce sparsity in regression coefficient and the Ising prior is used to capture structural relationship among predictors. The ICM/M algorithm is also introduced to cycle through each coordinate of the parameters in the model. This paper intends to extend Empirical Bayes variable selection for GLMs. We here focus on two important models arised in analyzing genomic data including the binary logistic and Cox's proportional hazards models. A main challenge of such an extension has emerged as how to generalize the framework from linear to nonlinear models. Here we borrow the idea of IRLS like in fitting classical GLMs to the ICM/M algorithm. This simple-to-implement algorithm achieves fast computation even in high-dimensional data analysis. The ICM/M algorithm for both linear and nonlinear models is implemented in the \texttt{icmm} R package.

The lasso is the most widely used method in fitting high-dimensional GLMs. Its popularity stems from its computational feasibility and availability in various packages. Our results show that ICM/M with lasso fits as initial values improves significantly over lasso in terms of number of false positives for both logistic and Cox's proportional hazards model. It can reduce the model size from lasso substantially by thresholding property of the ICM/M algorithm. For binary logistic model, we observed that the predictive ability of our proposed method was also better than lasso. For Cox's model, ICM/M outperformed lasso and adaptive lasso in terms of the ability to select correct variables into the model. ICM/M is also compuational feasible even for large $p$.

In biomedical field, most researchers would like to integrate expert knowledge such as gene regulartory networks in data analysis. While lasso and adaptive lasso do not utilize any prior information on strutural information among predictors, an important feature of ICM/M is that it can also incorporate more complicated prior easily. Here we demonstrated how to integrate a Markov random field representing relationships among predictors through an Ising model for binary logistic and Cox's models. Simulation studies and real data show a considerably improvement of ICM/M from lasso.

Another advantage of our proposed methodology over lasso and adaptive lasso is that our method does not only provide regression coefficients but also a local posterior probability. A local posterior probability for each predictor is proposed to quantify the importance of variable in linear model (\cite{Pungpapong2015}). For Bayesian variable selction for GLMs, the local posterior probability is estimated using the pseudodata in the last iteration of IRLS. FDR in Bayesian scheme can also be computed and employed to select the final set of important variables. Therefore, our method provides flexibility to select varibles while controlling FDR at a desired level rather than relying on only regression coefficients.

Finally, we would like to note that our method seems to work better with survival data than binary data in terms of false negative rate. This might be due to the fact that binary data is less informative than continuous data even though the censoring rate is quite high (50\%).

\section*{Acknowledgement}


\section*{References}

\bibliography{ICMM_GLMs}

\section*{Supplementary Table}
\begin{center} \footnotesize
	\begin{longtable}{l | c c c | c c c }
		\caption*{Table A: Results from PD data analysis: details on identified genes (Bolded genes were identified by all methods).} \label{table:PD-identifiedgenesallmethods} \\
		\hline
		\multirow{2}{*}{Gene (Location)} & \multicolumn{3}{|c|}{Number of identified SNPs} & \multicolumn{3}{|c}{p-value} \\
		&  Lasso & ALasso & ICM/M & Lasso & ALasso & ICM/M \\
		\hline
		\endfirsthead

		\caption*{Table A: Results from PD data analysis: details on identified genes (Bolded genes were identified by all methods) (continued).} \\
		\hline
		\multirow{2}{*}{Gene (Location)} & \multicolumn{3}{|c|}{Number of identified SNPs} & \multicolumn{3}{|c}{p-value} \\
		&  Lasso & ALasso & ICM/M & Lasso & ALasso & ICM/M \\
		\hline
		\endhead
		
		\hline
		\multicolumn{7}{r}{\emph{Continue on next page}}
		\endfoot
		
		\hline
		\endlastfoot
		
		ATP5F1 (1p13.2) & 1 & 0 & 1 & 0.0088 & 1 &  1.01e-6 \\
		NDUFS2 (1q23) & 0 & 0 & 1 & 1 & 1 & 0.0010 \\
		\textbf{PIK3CD} (1p36.2) & 1 & 1 & 1 & 5.93e-8 & 0.0037 & 1.14e-8 \\
		CDC20 (1p34.1) & 1 & 0 & 0 & 0.0372 & 1 & 1 \\
		AKT3 (1q44) & 0 & 1 & 0 & 1 & 0.0113 & 1 \\
		LCLAT1 (2p23.1) & 1 & 0 & 0 & 0.0150 & 1 & 1 \\
		BIRC6 (2p22.3) & 2 & 0 & 1 & 5.23e-8 & 1 & 2.13e-9 \\
		\textbf{TRIP12} (2q36.3) & 1 & 1 & 1 & 0.0352 & 0.0638 & 4.51e-5 \\
		\textbf{NDUFB4} (3q13.33) & 159 & 141 & 62 & 1.16e-70 & 3.2e-58 & 8.69e-66 \\	\textbf{HGD} (3q13.33) & 159 & 141 & 62 & 1.16e-70 & 3.2e-58 & 8.69e-66 \\
		PRKAR2A (3p21.3-p21.2) & 0 & 0 & 1 & 1 & 1 & 0.9480 \\
		IL1RAP (3q28) & 2 & 0 &1 & 0.0021 & 1 & 3.17e-5 \\
		COX7B2 (4p12) & 1 & 1 & 0 & 0.153 & 0.0091 & 1 \\
		UCHL1 (4p14) & 0 & 1 & 1 & 1 & 0.0317 & 0.0146 \\
		\textbf{SNCA} (4q21) & 1 & 1 & 1 & 0.0043 & 0.0029 & 0.0020 \\
		\textbf{ATP6V1G2} (6p21.3) & 48 & 38 & 17 & 2.41e-28 & 5.99e-24 & 1.26e-38 \\
		\textbf{TNF} (6p21.3) & 48 & 38 & 17 & 2.41e-28 & 5.99e-24 & 1.26e-38 \\
		\textbf{PARK2} (6q25.2-q27) & 5 & 5 & 3 & 3.29e-8 & 2.42e-5 & 1.34e-17 \\
		DDC (7p12.2) & 0 & 2 & 0 & 1 & 0.8540 & 1 \\
		UBE2D4 (7p13) & 1 & 1 & 0 & 0.0786 & 0.0049 & 1 \\
		PRKAR2B (7q22) & 1 & 0 & 1 & 6.00e-4 & 1 & 7.01e-07 \\
		TNPO3 (7q32.1) & 0 & 1 & 0 & 1 & 0.0950 & 1 \\
		\textbf{ATP6V0A4} (7q34) & 2 & 1 & 1 & 0.0098 & 0.0934 & 0.0002 \\
		TCEB1 (8q21.11) & 0 & 1 & 0 & 1 & 0.0149 & 1 \\
		TNFRSF10B (8p22-p21) & 1 & 0 & 1 & 0.0205 & 1 & 0.0183 \\
		TYRP1 (9p23) & 1 & 1 & 0 & 0.1880 & 0.0062 & 1 \\
		\textbf{CUL2} (10p11.21) & 1 & 1 & 1 & 0.0339 & 0.0034 & 0.0005 \\
		ATP5C1 (10p15.1) & 0 & 1 & 0 & 1 & 0.1770 & 1 \\
		FAS (10q24.1) & 0 & 0 & 1 & 1 & 1 & 0.0044 \\
		LHPP (10q26.13) & 0 & 1 & 0 & 1 & 0.4230 & 1 \\
		\textbf{PRMT3} (11p15.1) & 1 & 2 & 1 & 3.97e-5 & 0.0001 & 2.58e-7 \\
		PSMD13 (11p15.5) & 0 & 1 & 0 & 1 & 0.0068 & 1 \\
		TYR (11q14-q21) & 0 & 1 & 0 & 1 & 0.1320 & 1 \\
		NDUFA12 (12q22) & 1 & 0 & 0 & 0.3050 & 1 & 1 \\
		UBE2N (12q22) & 0 & 1 & 0 & 1 & 0.0128 & 1 \\
		UBE3A (15q11.2) & 0 & 1 & 0 & 1 & 0.0214 & 1 \\
		UBE2Q2 (15q24.2) & 0 & 1 & 0 & 1 & 0.0205 & 1 \\
		DET1 (15q25.3) & 1 & 2 & 0 & 0.1450 & 0.0279  & 1 \\
		NDUFB10 (16p13.3) & 0 & 1 & 0 & 1 & 0.0270  & 1 \\
		\textbf{MAPT} (17q21.1) & 107 & 51 & 72 & 7.11e-52 & 2.78e-28 & 6.17e-66 \\
		\textbf{ANAPC11} (17q25.3) & 107 & 51 & 72 & 7.11e-52 & 2.78e-28 & 6.17e-66 \\
		NEDD4L (18q21) & 0 & 1 & 0 & 1 & 0.9180 & 1 \\
		ECH1 (19q13.1) & 0 & 1 & 0 & 1 & 0.0252 & 1 \\
		AKT2 (19q13.1-q13.2) & 0 & 1 & 0 & 1 & 0.6110 & 1 \\
		CBLC (19q13.2) & 0 & 1 & 0 & 1 & 0.0813 & 1 \\
		NDUFA11 (19p13.3) & 1 & 0 & 1 & 0.0118 & 1 & 8.15e-5 \\
		ATP4A (19q13.1) & 1 & 1 & 0 & 0.0682 & 0.0973 & 1 \\
		UBE2G2 (21q22.3) & 2 & 0 & 0 & 0.3360 & 1 & 1 \\
		ATP5O (21q22.11) & 0 & 0 & 1 & 1 & 1 & 7.98e-9 \\
	\end{longtable}
\end{center}

\end{document}